\documentclass[useAMS,graphics,usenatbib]{mn2e}
\usepackage{amssymb,color}
\usepackage{epsfig}

\newcommand{\traphic}{\textsc{TRAPHIC}}

\newcommand{\Myr}{~\mbox{Myr}}

\newcommand{\eV}{~\mbox{eV}}

\newcommand{\cmsq}{~\mbox{cm}^{2}}

\newcommand{\K}{~\mbox{K}}

\newcommand{\mr}[1]{{\rm #1}}     

\newcommand{\nh}{n_\mr{H}}

\newcommand{\apj}{ApJ}

\newcommand{\mnras}{MNRAS}

\newcommand{\aap}{A\&A}

\title[Recombination radiation in \traphic]{The effect of recombination radiation on the temperature and ionization state of partially ionized gas}

\author[M. Rai\v{c}evi\'{c} et al.]
{Milan Rai\v{c}evi\'{c}$^{1}$\thanks{E-mail:milan.raicevic@strw.leidenuniv.nl}, Andreas H. Pawlik$^2$, Joop Schaye$^1$, Alireza Rahmati$^1$\\
  $^1$Leiden Observatory, Leiden University, P.O. Box 9513, 2300 RA Leiden, The Netherlands\\
  $^2$Max-Planck Institute for Astrophysics, Karl-Schwarzschild-Strasse 1, 85748 Garching, Germany}

\begin{document}

\date{}

\pagerange{\pageref{firstpage}--\pageref{lastpage}} \pubyear{2010}

\maketitle

\label{firstpage}

\begin{abstract}

A substantial fraction of all ionizing photons originate from
radiative recombinations. However, in radiative transfer calculations
this recombination radiation is often assumed to be absorbed
``on-the-spot'' because for most methods the computational cost
associated with the inclusion of gas elements as sources is
prohibitive.  We present a new, CPU and memory efficient
implementation for the transport of ionizing recombination radiation
in the TRAPHIC radiative transfer scheme. TRAPHIC solves the radiative
transfer equation by tracing photon packets at the speed of light and
in a photon-conserving manner in spatially adaptive smoothed particle
hydrodynamics simulations. Our new implementation uses existing
features of the TRAPHIC scheme to add recombination radiation at no
additional cost in the limit in which the fraction of the simulation
box filled with radiation approaches 1. We test the implementation by
simulating an HII region in photoionization equilibrium and comparing
to reference solutions presented in the literature, finding excellent
agreement. We apply our implementation to discuss the evolution of the
HII region to equilibrium. We show that the widely used case A and B
approximations yield accurate ionization profiles only near the source
and near the ionization front, respectively. We also discuss the
impact of recombination radiation on the geometry of shadows behind
optically thick absorbers. We demonstrate that the shadow region may
be completely ionized by the diffuse recombination radiation field and
discuss the important role of heating by recombination radiation in
the shadow region.

\end{abstract}


\section{Introduction}

Many astrophysical problems require a detailed treatment of 
hydrogen-ionizing radiative transfer (hereafter RT). Examples include the
evolution of ionized regions during the epoch of reionization
\citep[e.g.,][]{barkana2001, bolton2004, ciardi2005, furlanetto2006, mellemab2006, iliev2007, trac2007, mcquinn2007,
  wise2008, aubert2010, zaroubi2013, norman2013}, the transport of stellar radiation through the
(dusty) inter-stellar medium \citep[e.g.,][]{ferland1998,
  devriendt1999, groves2008, jonsson2010} and its impact on the distribution and morphology 
of the interstellar gas \citep[e.g.,][]{mellema2006, susa2008, gritschneder2010, walch2013}, photoevaporation feedback
in galaxy formation \citep[e.g.,][]{shapiro1994, gnedin2000, dijkstra2004,
  okamoto2008, pawlik2009, petkova2011, finlator2011, hasegawa2013}, the properties of neutral
hydrogen absorbers exposed to the extra-galactic UV background
\citep[e.g.,][]{miralda-escude2003, mcquinn2011, altay2011, yajima2012, altay2013,
  rahmati2013a, rahmati2013b}, and others. However, radiative transfer is an
expensive operation, especially because most
methods used to transport ionizing radiation in simulations require a
computational cost that scales with the number of sources (see, e.g., 
the review of reionization simulation methods 
by \citealp{trac2009}, or \citealp{mackey2012}). This renders large simulations containing many
sources computationally challenging.

Apart from discrete sources of radiation such as stars or accreting
black holes, about 40\% of the recombinations of hydrogen are
accompanied by the emission of ionizing photons, assuming gas of
temperature $T=10^4 \K$ as is typical of photoionized regions. The presence of this diffuse recombination
radiation component greatly increases the number of sources, thus
increasing the numerical challenge. To reduce the computational cost
associated with recombination radiation, the majority of works in the
literature employ the ``on-the-spot'' approximation
\citep[e.g.,][]{spitzer1978, osterbrock1989}, which assumes that the
ionizing photons produced in recombinations are absorbed locally, in
the immediate surroundings of the recombining parcel of gas. In this
approximation, recombination radiation is accounted for by a simple
change in the recombination coefficient.

However, the on-the-spot approximation ignores RT effects that may
be important. \cite{ritzerveld2005} showed, using an outward-only
analytic approximation, that the intensity of the diffuse recombination
radiation should dominate over that of the direct stellar radiation
near the ionization fronts (henceforth I-fronts) of HII regions in
ionization equilibrium, finding a  diffuse-to-source intensity 
ratio substantially larger than implied by the on-the-spot
approximation (for an early discussion see \citealp{rubin1968}). \cite{williams2009} revisited this issue by carrying
out accurate numerical RT simulations of HII
regions in ionization equilibrium. They confirmed the picture provided
by \cite{ritzerveld2005} qualitatively, but showed that the
outward-only approximation underestimates the contribution of
recombination radiation to the total intensity near the source and
overestimates it near the I-front, and also discussed the dependence
of the diffuse-to-source intensity ratio on the radiation spectrum. Similar
recent investigations into the impact of ionization by recombination
radiation on the intensity distribution or ionization balance inside HII
regions include \cite{hasegawa2010}, \cite{cantalupo2011} and \cite{friedrich2012}.
\par
Recombination radiation may strongly affect the morphology of shadows
cast by optically thick absorbers.  \cite{canto1998}, using an 
approximate analytic RT model of an optically thick clump illuminated
by a central source in a stellar HII region, showed that recombination
radiation may highly ionize and photoheat the gas inside the shadow. In
their simulations, the dynamical reaction induced by the difference in
the temperatures of the gas in- and outside the shadow increases the
gas densities until pressure equilibrium is reached, possibly rendering 
the shadow observable in H$\alpha$ against the HII region. 
Shadowing by optically thick clumps was also investigated
in \cite{razoumov1999} using three-dimensional RT and in
\cite{pavlakis2001} using an approximate, inward-only RT scheme coupled
to a two-dimensional hydrodynamical simulation, and
both these works confirmed that recombination radiation can strongly
impact the shadow morphology (see also \citealp{hasegawa2010}; 
\citealp{rosdahl2013}). On the
other hand, \cite{raga2009}, who modeled the ``finger'' structures in
the Carina Nebula as overlapping shadow regions behind multiple clumps
of gas using three-dimensional radiation-hydrodynamical simulations, 
found no significant impact of recombination radiation on the 
morphology of the simulated nebula.
\par
Other investigations into the role of ionizing recombination radiation
on small-scale gaseous structure include \cite{williams1999}, who
discussed the potentially inhibiting impact of recombination radiation
on the growth of I-front instabilities (see also
\citealp{whalen2008}). \cite{motoyama2007} simulated the compression
of gas clumps exposed to diffuse radiation emitted by the surrounding
gas.  Adopting a directed, non-isotropic approximation for the RT of diffuse
radiation, they showed that recombination photons can photoheat and
compress the gas in the clump, increasing the rates at which clumps
accrete, possibly implying an increase in the luminosities of the
emerging protostars. The effects of recombination radiation on star
formation triggered by radiation-driven implosion of gas clumps were
also investigated in the three-dimensional radiation-hydrodynamics
simulations of \cite{haworth2012}. \cite{ercolano2011} approximated
the effects of recombination radiation in simulations of turbulent
interstellar gas photoionized by a nearby star, finding less dense
and less coherent structures and a small reduction in the efficiency
with which turbulence is driven with respect to simulations adopting
the on-the-spot approximation. Recombination radiation is also
generated in shocks accompanying accretion of gas onto galaxies or
protostellar jets, where it may affect the evolution of the unshocked
gas ahead of the shock (e.g., \citealp{kang1992}; \citealp{raga1999};
see also \citealp{dopita2011}; \citealp{wyithe2011}).

The role of diffuse radiation in the epoch of reionization is largely
unexplored. \cite{ciardi2001} explicitly followed recombination
radiation in their Monte Carlo RT simulations of reionization, but did
not discuss its impact. \cite{inoue2010} used an analytic model to
show that accounting for the RT of recombination radiation can impact
both the escape of ionizing photons from high-z galaxies, and the
average energy carried by the escaping radiation. More importantly,
they showed that diffuse nebular ionization leaves a clear Lyman
``bump'' signature in the spectral energy distributions of galaxies
that can be used to characterize the properties of stellar populations in 
star-forming galaxies.

\citet{rahmati2013a, rahmati2013b} used the RT technique described in
this paper to investigate how ionizing radiation from stars in
galaxies and from the UV background shapes the distribution of neutral
hydrogen absorbers in and near galaxies in the post-reionization
era. They found that the recombination radiation intensity dominates over
the UV background intensity in galaxies near densities $\rm n_{\rm H} \sim
10^{-2} cm^{-3}$ above which hydrogen starts to self-shield, helping
the UV background photons to penetrate to somewhat higher
densities. Even in the presence of ionization by local stellar
sources, which typically dominate the intensity in the interstellar gas,
recombination radiation may dominate the intensity near the densities at
which the gas self-shields from the UV background.

In this paper, we present a new implementation of recombination
radiation in the RT scheme \traphic\ \citep{pawlik2008,
  pawlik2011}. The key idea of this implementation is to reuse the
original features of the \traphic\ code to add the treatment of
recombination radiation at almost no additional computational cost in
the limit when the maximum number\footnote{The number of photon
  packets present in the simulation box has a maximum value in
  TRAPHIC, which depends on the angular resolution at which photon
  packets are merged. See Section \ref{sect:traphic} for more
  details.}  of photon packets is present in the box. This is the
relevant limit for the intended use of TRAPHIC in, e.g., simulations
of galaxy formation close to and after the end of reionization. Away
from this limit, the new implementation can still be significantly
more computationally efficient than the original implementation described
in \citet[their Section 4.3.2]{pawlik2008}, in which the recombination 
radiation was emitted using the same framework as used for the emission 
of stellar radiation.

The paper is structured as follows. In Section \ref{sect:traphic}, we
present a brief overview of the \traphic\ method. In Section
\ref{sect:recrad_method}, we describe our new implementation of
recombination radiation. In Section \ref{sect:stromgren}, we apply our
implementation in Test 1 of \cite{iliev2006} and we compare our
results to the analytic solution of \cite{ritzerveld2005} and the
numerical solution by \cite{williams2009}. In Section
\ref{sect:shadow}, we use our method to investigate the role of
recombination radiation in the shadowing by an absorber near a UV
radiation source. This problem is similar in spirit to that studied by, e.g.,
\cite{canto1998}, \cite{razoumov1999}, and \cite{rosdahl2013}. 
Finally, we conclude in Section \ref{sect:conclude}.

\section{Method}

In this section, we provide a concise overview of the \traphic\ method
and the implementation of recombination radiation.

\subsection{\traphic: TRAnsport of PHotons In Cones}
\label{sect:traphic}

\traphic\ is a photon packet-based multi-frequency RT scheme designed
to transport radiation in smoothed particle hydrodynamics (hereafter
SPH) simulations. It is implemented in GADGET \cite[last described
  in][]{springel2005}. The scheme itself was presented in
\cite{pawlik2008, pawlik2011}, and later applied in post-processing by
\cite{rahmati2013a,rahmati2013b} and in radiation-hydrodynamical
simulations by \cite{pawlik2013}. \traphic\ solves the time-dependent
RT equation in an explicitly photon-conserving manner by tracing
photon packets at the speed of light from the sources through the simulation
box. Packets are traced directly on the SPH particles, thus utilizing
the full dynamic range of the density field. A key feature of the
method is the directional merging of photon packets inside cones,
which limits the maximum number of photon packets traced and renders
the computational cost of TRAPHIC independent of the number of
radiation sources.

The RT calculation proceeds in 2 steps: the radiation transport step
and the chemistry step. During the RT step of duration $\Delta t$,
photons are propagated and absorbed by the gas. During the chemistry
step, the absorptions are used to update the species fractions over
the same duration $\Delta t$. The associated photoheating and radiative cooling 
is also taken into account. Particles that are sources of radiation emit new
photon packets isotropically in $N_{\rm EC}$ directions, where $N_{\rm
  EC}$, the ``number of emission cones'', is a parameter chosen to
obtain converged results. The emission is done in steps\footnote{This decoupling of the radiative transfer
and emission steps is a new feature that has not been described in \citealp{pawlik2008}. It enables additional 
control of the angular sampling of the volume around sources independent of the size of the RT time step 
$\Delta t$ and the number of emission cones $N_{\rm EC}$.} of $\Delta
t_{\rm em} \leq \Delta t$. Before each emission, the isotropic
directions are randomly rotated to increase the fraction of the
simulation box sampled with photons.

Each source in TRAPHIC can have an arbitrary spectrum, defined by a
number of bins at discrete frequencies, $N_{\rm freq}$. Each photon
packet carries photons in a single frequency bin and hence the total
number of packets emitted by a source in an emission time step is
$N_{\rm EC} \times N_{\rm freq}$. The spectral shape within each bin
is encoded in the form of a grey absorption cross-section computed by
averaging the frequency-dependent absorption cross-section weighted by
the source spectrum. This so-called grey approximation is accurate in
the optically thin limit, when spectral hardening within each bin is
negligible \cite[e.g.,][]{osterbrock1989}.

The newly emitted photons, together with any photons already present
in the box, are then propagated forward. In each forward step, photon
packets search for neighbours $j$ within the smoothing volume of SPH
particle $i$ they are attached to. A given photon packet is then
distributed among the neighbours found in the regular cone with
opening solid angle of $4 \pi / N_{\rm TC}$ centred on the packet's
propagation direction. The parameter $N_{\rm TC}$, called the ``number
of transmission cones'', sets the angular resolution of the
transport. The distance $l_{ij}$, which is the distance to neighbour
$j$ projected onto the direction of propagation, is used to compute
the optical depth of the absorbing species $k$ (where $k$ is HI, HeI
or HeII) using $\tau_{ij, \nu}^{k} = \langle \sigma_{k} \rangle_{\nu}~n_{j}^{k} l_{ij}$,
where $\langle \sigma_{k}\rangle_{\nu}$ is the grey cross-section for absorption of the
photon packet in frequency bin $\nu$ by species $k$ and $n_{j}^{k}$
the number density. 
\par
The optical depth gives the number of photons
absorbed by the gas along the travel length,
$\mathcal{N}_{\rm{abs},\nu} = \mathcal{N}_{\rm{in},\nu}[1-
  \exp(-\tau_{ij,\nu})]$, where $\mathcal{N}_{\rm{in},\nu}$ is the
number of photons in the packet before the transport step and
$\tau_{ij, \nu} = \sum_{k} \tau_{ij, \nu}^{k}$. Within each RT time step
$\Delta t$, photon packets are moved until they reach the distance set
by the speed of light, $c \Delta t$, i.e., they generally experience
several forward propagations and move across multiple inter-particle
distances in one transport step. The distance covered is measured by a 
{\it clock} associated with each photon packet, and which is initialized upon emission using the
time at the beginning of the current RT time step. Corrections due to 
relativistic effects, which we ignore, are negligible in the applications discussed in this work.
\par
It is possible that a photon packet does not find any neighbours in
its transmission cone. In that case, a so-called virtual particle is
created randomly within the volume of the transmission cone. The properties of the
virtual particle are obtained by SPH interpolation from its SPH
neighbours. Its role is to enable the transport of photons in
directions without SPH neighbours, after which it is deleted. All
absorptions that happen during the transport of photon packets to a
virtual particle are distributed among its SPH neighbours and
contribute to their ionization and thermal evolution.

\begin{figure*}
  \begin{center}
    \includegraphics[width=0.9\textwidth,clip=true, trim=10 10 10 10,
      keepaspectratio=true]{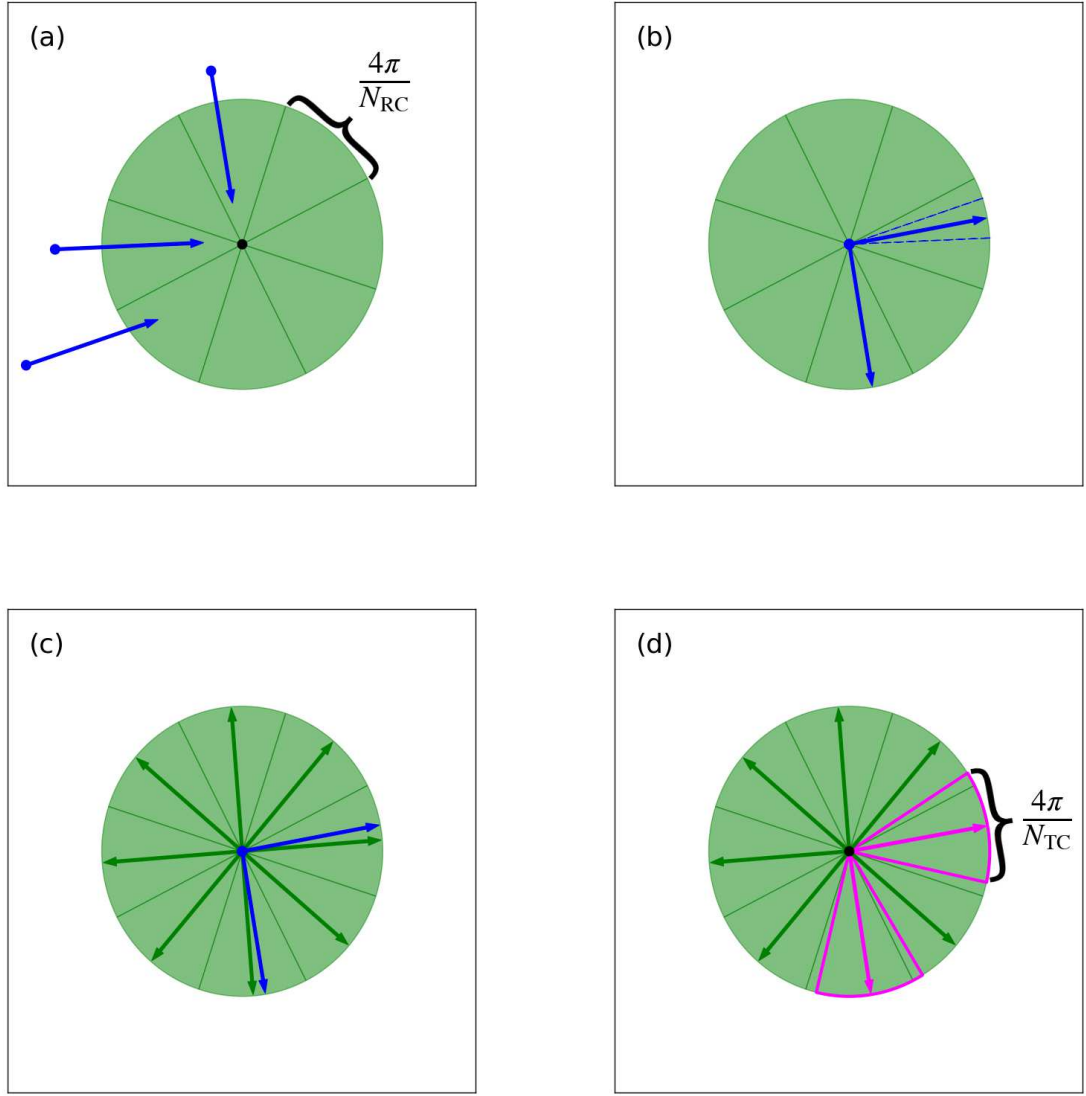}
  \end{center}
  \caption{The emission of recombination photons in the new
    implementation discussed here. The diagrams are a two-dimensional
    representation of the three-dimensional, $4 \pi$ solid angle around particles. {\it (a)} An SPH particle (black
    dot) has a set of isotropically directed reception cones attached to it (green
    cones). Photon packets (blue arrows) arriving at this particle are
    grouped according to their incoming directions using the reception
    cones. {\it (b)} All packets whose propagation directions fall in the same reception cone
    are merged into one packet. The direction of the merged packet is
    set to the luminosity-weighted average of the parent directions. Blue
    dashed lines show the direction of parent photons before merging,
    for reference. {\it (c)} The isotropic directions that define the
    reception cones are used to emit recombination radiation from the
    particle (green arrows). {\it (d)} The newly generated
    recombination radiation packets are instantly merged with existing
    photons present at the SPH particle (magenta arrows). Note that,
    for visualization, we assumed that the luminosity of pre-existing
    photons (blue arrows) is much higher than that of the newly
    emitted ones (green arrows), causing the directions of merged
    packets (magenta arrows) to nearly coincide with the blue arrows; we furthermore 
    chose $N_{\rm TC} = N_{\rm RC}$, making the indicated opening
    angles of the cones in panels {\it (a)} and {\it (d)} equal in size. }
  \label{fig:merge_recrad_emit_diagram}
\end{figure*}

Multiple photon packets can arrive at the same SPH particle in one
transport step. Then, packets are merged according to their incoming
direction, as illustrated in Figure
\ref{fig:merge_recrad_emit_diagram}. This is achieved by defining a
number of isotropic directions, $N_{\rm RC}$, that define directional
bins called reception cones. These reception bins are shown as green
cones in Figure \ref{fig:merge_recrad_emit_diagram}. Each photon
packet is sorted in one of these bins according to its direction (top
left panel). Photon packets whose propagation directions fall into the same bin 
are combined into a single new photon packet. The propagation direction of this new, merged 
photon packet is set to a luminosity-weighted average of the propagation directions of the progenitor packets (top
right panel). The reception directions are frequently rotated to avoid
geometrical artifacts. The merging of photons is done separately for
each frequency bin. This prevents the mixing of photons from different
bins, which enables an accurate tracking of the contribution of
individual spectral components to the total intensity. Thanks to this
merging procedure, the number of photon packets in the
computational volume is at most $N_{\rm RC} \times N_{\rm SPH} \times
N_{\rm freq}$, where $N_{\rm SPH}$ is the number of SPH particles. The 
computational cost of simulations with TRAPHIC is thus
independent of the number of sources.
\par
Once the transport step has completed, the ionization state and
thermal energy are updated. The total number of absorptions at each
SPH particle is used to estimate the rate of ionizations by photons in frequency bin $\nu$ for species
$k$, $\Gamma_{\gamma k, \nu} = \frac{\mathcal{N}_{\rm{abs,
      tot},k, \nu}}{\mathcal{N}_{k} \Delta t}$, where
$\mathcal{N}_{\rm{abs, tot},k, \nu}$ is the sum of all photon packets in frequency bin $\nu$
absorbed by species $k$ within the time step $\Delta t$ and
$\mathcal{N}_{k}$ is the number of atoms of species $k$ in the
SPH particle. The total photoionization rate is then the sum of the contributions from all 
frequency bins. Currently implemented species are HI, HeI and HeII. The
associated photoheating rate due to ionizations of each species by photons in frequency bin
$\nu$ is $\mathcal{H}_{\gamma k, \nu} = \Gamma_{\gamma k, \nu} \langle 
\epsilon_{k} \rangle_{\nu}$, where $\langle \epsilon_{k} \rangle_{\nu}$ is the grey excess
heating energy per ionization of species $k$ in frequency bin $\nu$, computed as described in
\citet[here we adopt their optically thin limit]{pawlik2011}. 
Once the photoionization and photoheating rates have been obtained,
they are used to compute the change in the ionization state and
temperature during the time step $\Delta t$. For details on our method
for solving the chemistry equations, see \cite{pawlik2011}. 
\par
We use the fits to the recombination rate coefficients by \cite{hui1997}. We
account for collisional ionization and radiative cooling by
collisional ionization/excitation, emission of free-free and
recombination radiation and Compton radiation off the CMB using the
rates described in \citet[their Table 1]{pawlik2011}. Unless mentioned
otherwise, we adopt Case B recombination and cooling rates in
simulations employing the on-the-spot approximation, and Case A
recombination and cooling rates when recombination radiation is
explicitly followed using RT. Expressions for the frequency-dependent
photoionization cross-sections are taken from
\cite{verner1996}. Photoionizations are computed in the nebular approximation,
i.e., assuming that the electrons in the hydrogen atoms are in the
ground state when photoionizations occur, and hence require 
photon energies of at least $13.6 \eV$ to be freed (e.g., \citealp{osterbrock1989}).
\par
Emitting a finite number of photon packets in a finite number of
directions may imprint noise in the ionization and temperature fields,
as known from simulations with classical Monte Carlo radiative
transfer codes. The merging of photon packets may amplify this
noise. In simulations with TRAPHIC, numerical noise caused by the
Monte Carlo nature of the emission of photons and by the merging is
controlled by numerical parameters whose values can be informed by
convergence tests. We have presented such tests in \cite{pawlik2008}, and we discuss convergence of our
results in the current simulations in Appendix A. In
radiation-hydrodynamical simulations, Monte Carlo noise may cause
artificial pressure differences that affect the dynamics of the
gas. Such differences are also controlled by the numerical parameters
and can be reduced to have a negligible effect (see, e.g., the
discussion in \citealp{iliev2009}). However, in this work we 
consider only the transport of photons on static density fields.
\par

\subsection{Implementation of recombination radiation}
\label{sect:recrad_method}

In this section, we introduce the new implementation of hydrogen
recombination radiation in TRAPHIC and discuss the differences
relative to the original implementation presented in \citet[their Section
  4.3.2]{pawlik2008}. Note that we had not previously described the
results of our tests of the transport of recombination radiation in
TRAPHIC, but we already included the transport of diffuse photons (using
the new implementation described here) in some of our previous works
(\citealp{rahmati2013a,rahmati2013b}).
\par
Recombination is the inverse process of photoionization, during which
a free electron is captured by an ion into an atomic energy level
$l$. This is followed by the emission of a recombination photon with
energy corresponding to the difference between the energy of the
atomic level and that of the electron before recombination. The energy
of the emitted photon can in some cases be sufficient to cause another
photoionization event. In this paper, we focus on accounting for the
ionizing recombination radiation in a hydrogen-only medium. In
general, the balance between photoionizations and recombinations is
governed by the Milne relations \citep[e.g.,][]{rybicki1986}. These
relations can be used to compute the probability of electron capture
to an atomic energy level $l$, taking into consideration the thermal
distribution of the electrons. The total temperature-dependent rate of
recombinations is obtained by integrating over captures to {\it all}
atomic energy levels. This rate is often referred to in the literature
as the {\it case A} recombination rate, $\alpha_{\rm A}(T)$
\citep[e.g.][]{spitzer1978}. The recombinations to the ground level
$l=1$, occurring at a rate $\alpha_{\rm 1}(T)$, are guaranteed to
generate hydrogen-ionizing photons.
\par
\begin{figure*}
  \begin{center}
    \includegraphics[width=0.9\textwidth,clip=true, trim=10 10 10 10,
      keepaspectratio=true]{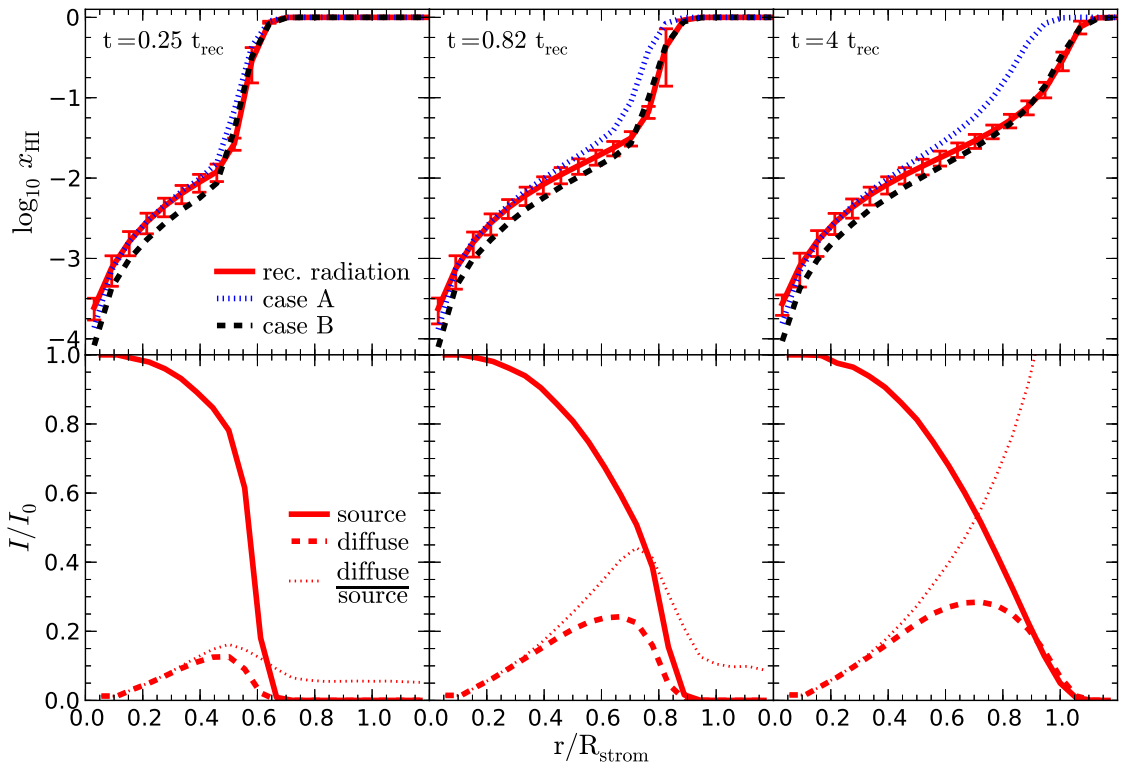}
  \end{center}
  \caption{Evolution of the HII region in Test 1 of the Code
    Comparison project (Iliev et al. 2006), comparing different
    treatments of recombination radiation. {\it Top:} The radial
    profile of the neutral fraction $x_{\rm HI}$ at three different
    times, quoted in units of the recombination time, $t_{\rm rec}=122.4$ Myr. The solution obtained using accurate RT of recombination photons 
    (red solid curve) is compared to the solutions obtained
    using case A (blue dotted curve) and case B (black dashed curve)
    recombination coefficients and ignoring the transport of
    recombination radiation. The RT solution accurately
    interpolates between the optically thin and thick regimes. {\it
      Bottom:} The local radiation intensity $I$ multiplied by $4 \pi r^2$, measured 
    in thin spherical shells as described in the text, in units of $I_{0}$, 
    the number of photons emitted per unit time by the source. Contributions to the intensity from the source (red solid
    curve) and from diffuse recombination radiation (red dashed curve)
    are shown separately, together with the diffuse-to-source intensity
    ratio (thin red dotted curve). When the HII region has reached
    the equilibrium state, the diffuse component dominates the intensity
    near the I-front (bottom right panel). }
  \label{fig1}
\end{figure*}

In a RT simulation, explicitly following these photons generally
represents a significant addition to the computation cost, as each
recombining ion becomes a new isotropic source of radiation. To avoid
this additional cost, the ``on-the-spot'' approximation was introduced
\citep[e.g.][]{spitzer1978, osterbrock1989}, which assumes that the
optical depth of the region around the recombining gas is high enough
to absorb all the ionizing recombination photons locally. This is implemented
by excluding recombinations to the ground state from the total
recombination rate, introducing the {\it case B} recombination rate,
$\alpha_{\rm B} = \alpha_{\rm A}-\alpha_{\rm 1}$. Both case A and case
B recombination rates can be appropriate substitutes for the explicit
RT of recombination radiation in some regimes, but not in general, as
we will demonstrate in the next section (Figure \ref{fig1}).
\par
In the original implementation described in \cite{pawlik2008}, the
emission of recombination radiation was done using the same procedure
as used for the emission of photons from stars. That is, an additional
set of $N_{\rm EC}$ isotropic emission cones was defined for each SPH
particle. The $N_{\rm EC}$ photon packets emitted in these cones were
stored on the emitter particle {\it separately} from any other packets
already present. In order to avoid the additional memory and CPU cost
associated with introducing additional emission cones to SPH
particles, our new implementation {\it reuses} the $N_{\rm RC}$
reception cones, which are already in place, to emit
photons\footnote{The same procedure was used by \cite{rahmati2013b} to
  emit radiation from star-forming SPH particles.}. This is done by
adding the recombination radiation isotropically into the reception
cones at the end of each chemistry step $\Delta t$ (once the
recombination radiation luminosity for that step is known), instantly
merging with any photons packets already present (see Figure
\ref{fig:merge_recrad_emit_diagram}). The clocks used to propagate the
photon packets at the speed of light are initialized using the 
time at the beginning of the next RT time step. 
\par
The efficiency of the new implementation comes from the fact that
almost nothing was added to the TRAPHIC scheme to accomplish the
emission of recombination radiation, as the employed reception cones
already exist and the memory to store them has already been
allocated. The emission in reception cones differs from the emission
using emission cones in two respects: the solid angle subtended by the
cones ($4\pi / N_{\rm RC}$ instead of $4\pi / N_{\rm EC}$) and the
frequency with which the cones are rotated. Moreover, in the current work, 
the emission of recombination photons is done on the chemistry update time step,
$\Delta t$. This contrasts with the emission of stellar photons and
the emission of recombination radiation in the original implementation,
which was done on time steps of size $\Delta t_{\rm em} \leq \Delta t$. The
recombination radiation is therefore typically emitted in fewer directions in
the new implementation. The coarser angular sampling is, however,
offset by the large number of sources of recombination radiation (all
SPH particles are potential sources of recombination
radiation). Indeed, our tests show fast convergence even for small
values of $N_{\rm RC}$ (Appendix A). Comparing simulations using the
new and original implementations yields nearly identical results.
\par
A further improvement in memory efficiency can be made by using the
same frequency bins for transporting both stellar and recombination
radiation. The HI ionizing recombination radiation spectrum is strongly peaked
at the Lyman continuum limit, i.e., at energies $\gtrsim 13.6\eV$ 
(e.g., \citealp{canto1998}). The spectrum of massive stars 
also features a significant number of photons close to the
Lyman limit, thus a frequency bin around $13.6 \eV$ is necessary to
accurately represent it, possibly complemented by additional frequency bins at higher energies. 
If that frequency bin is also used to transport recombination radiation, the addition of
recombination radiation implies practically no extra cost. However,
this is not the approach we take in this paper. It is only appropriate
if the stellar spectrum in the frequency bin and the spectrum of the
recombination radiation are similar. The general case would require to
average the cross-sections for the absorption of stellar and
recombination photons upon merging of photons in the same
frequency bin. In all tests presented in this paper, the recombination
radiation is transported in a single frequency bin physically separate from, though overlapping 
in frequency with, the bin used to transport direct stellar radiation. This allows us
to determine unambiguously the contribution of recombination radiation
to the total intensity.

\par

\par
In this work we present simulations of the transfer of recombination
radiation on static hydrogen-only density fields. We thus do not
consider the emission and absorption of helium recombination radiation
and the associated ionization and heating in this work (e.g.,
\citealp{cantalupo2011}; \citealp{friedrich2012}). In some
simulations, we also follow the associated photoheating of the
gas. The excess energy injected upon absorption of hydrogen
recombination photons available to heat the gas is set following
\cite{canto1998} to $\epsilon_{\rm HI, RR} = k T$, where $T$
is the temperature of the gas emitting recombination radiation. This
assumes that the spectrum of recombination radiation is proportional 
to $e^{-h \nu / kT}$. However, we ignore the spectral distribution of the recombination radiation 
when evaluating the grey cross-section for absorption by neutral hydrogen
atoms. Throughout this work, we adopt a grey photoionization cross-section for recombination radiation 
of $\langle \sigma_{\rm HI} \rangle =
6.3 \times 10^{-18} \cmsq$ appropriate for monochromatic photons of energy 
$13.6 \eV$. We opt for this approximation, which does not affect our general conclusions, 
because it facilitates the quantitative comparisons with previous works in Section~\ref{sect:stromgren}.

\section{The role of diffuse radiation in a Stromgren sphere}
\label{sect:stromgren}

In this section, we test our new recombination radiation
implementation by comparing results from a simulation of an HII region
with analytical and numerical results found in the
literature. Specifically, we compare the intensity distribution when the
HII region reaches equilibrium \citep[``Stromgren sphere'',
][]{stromgren1939} to the analytical solution of \cite{ritzerveld2005}
that assumes all transport is directed outwards, and to the numerical RT
equation solution of \cite{williams2009}. We begin by
discussing the evolution of the HII region as its I-front expands to
reach the Stromgren radius. We will compare the run with accurate RT
of recombination radiation to two runs employing two widely used
approximations, case A and B. 
\par
The simulation setup is as follows. We adopt the same parameters as in
Test 1 of the RT Code Comparison project \citep{iliev2006}. The
simulation box is cubic with linear size $\rm {\rm L}_{box} \rm =
13.2\, kpc$. It is filled with only hydrogen of uniform number density $ \nh \rm = 10^{-3} \, cm^{-3}$, 
which is initially neutral and has a constant temperature
$\rm T=10^4 \, K$ throughout the simulation. At the center of the box is a single star particle
emitting $\dot{N}_{\gamma} = 5 \times 10^{48} \rm \, ionizing \,
photons/s$. We transport the stellar radiation using a single
frequency bin, assuming monochromatic photons with energy $13.6 \eV$ and an 
absorption cross-section of $\langle \sigma_{\rm HI}
\rangle = 6.3 \times 10^{-18} \rm cm^{2}$. This differs from the test description in 
\cite{iliev2006}, who assumed
a black body spectrum with temperature $T=10^5 \, \rm K$, but facilitates the comparison 
with the equilibrium solutions presented in \cite{ritzerveld2005} and \cite{williams2009}. The hydrogen 
gas is discretised using $128^3$ SPH particles. At each emission time step of size $\Delta
t_{\rm em} = 0.01 \Myr$, the star emits $N_{\rm EC}=8$ photon
packets that are subsequently transported at angular resolution $N_{\rm TC}=8$. Photons are merged using
$N_{\rm RC} = 8$ reception cones per particle. Note that the parameter $N_{\rm RC}$ also sets the number of directions 
in which each gas particle emits recombination photons. The RT time step is set to
$\Delta t = 0.1 \Myr$. The latter time step is also used to emit
recombination photons. We discuss the convergence of our
simulation with respect to the spatial resolution, the angular
resolution and the time step in Appendix A.
\par
Figure \ref{fig1} shows the radial profiles of the HI fraction ({\it
  top}) and radiation intensity $I$ ({\it bottom}) during the evolution of the
HII region. The columns show our results at three characteristic
times, from the initial to the equilibrium state of the HII
region. The time is quoted in units of the recombination time, $t_{\rm
  rec} = (\alpha_{\rm B} \nh)^{-1} \approx 122.4 \rm \, Myr$, where
$\alpha_{\rm B}$ is the case B hydrogen recombination rate
coefficient. We compare the evolution of the HI fraction in the
simulation with explicit RT of recombination radiation (red solid
curve) to the corresponding evolution in simulations employing the
case A (blue dotted curve) and case B (black dashed curve)
approximations. 
\par
The run with the explicit RT of recombination radiation bridges the
results obtained using case A and B recombination rates. In the inner
part of the HII region, near the source, where the gas is already
highly ionized and has reached ionization equilibrium, the
recombination radiation RT result agrees with the case A one, while
case B yields a neutral fraction that is too low by a factor $\alpha_{\rm B} / \alpha_{\rm A} \approx 0.6$. 
The origin of the discrepancy with case B is the assumption that the
ionizing recombination photons are absorbed on the spot, which is not
appropriate in the highly ionized low-density gas that is optically
thin to ionizing radiation and from which the majority of the
recombination photons should escape. The latter is assumed in case A,
which therefore provides a good match to the RT result.
\par
On the other hand, around the I-front where the gas is optically thick
and the on-the-spot approximation appropriate, the RT result
approaches the case B one, while case A yields an I-front with a
smaller radius \citep[which we define as the radius at which the neutral
  fraction is $x_{\rm HI}=0.5$, following][]{iliev2006}. The reason
for the smaller radius of the I-front in the case A run is that case A
assumes that all recombination photons escape. While this assumption
is appropriate in optically thin regions, in optically thick regions
it leads to a violation of photon conservation and results in a
smaller I-front radius. The RT run correctly treats the ionization
state in both the optically thin and optically thick regions, meaning
it accurately interpolates between the two opacity regimes. A similar
discussion but focusing on the equilibrium solution was presented in
\cite{cantalupo2011}.
\par
The bottom panels in Fig. \ref{fig1} show the radial profiles of 
the intensity, separated in contributions from photons 
emitted by the source (black solid curves) and from recombinations (diffuse component, red dashed
curve). We measured the intensity by 
recording the total number of photons passing through each SPH particle
during a RT time step $\Delta t$ and dividing by this time step. Next, we sum these numbers 
in thin spherical shells centered on the central source. This yields
the total number of photons passing per unit time through a spherical surface with radius $r$, i.e.,  
$4 \pi r^2 I(r)$, which is the quantity discussed in \cite{williams2009} and
\cite{ritzerveld2005}. The profiles are normalized by dividing by the total number 
of photons emitted by the source per RT time step, $I_0$. The contribution of 
recombination radiation to the total intensity
increases with time. In the early stages of HII region growth, i.e.,
at times $t \ll t_{\rm rec}$ (bottom left panel), the contribution of
recombination radiation is sub-dominant, reaching at most $\sim 15 \%$
of the source intensity at any given radius. As the time approaches one
recombination time, the diffuse intensity contributes significantly to the
total, reaching up to $\sim 40\%$ of the source intensity near the
I-front. Finally, in equilibrium, i.e., at times $t \gg t_{\rm rec}$
(bottom right panel), the recombination radiation intensity starts to
dominate over the stellar intensity near the I-front. Note for comparison
that the Case B approximation yields a ratio of diffuse and source
intensity of about 0.6 independent of the distance from the source,
assuming that the cross-sections for absorption of diffuse and source
radiation are equal (\citealp{williams2009}). Our simulations show
that case B overestimates the diffuse intensity significantly at early
times or near the source, and underestimates it at late times near the
I-front.

\begin{figure}
  \begin{center}
    \includegraphics[width=0.48\textwidth,clip=true, trim=10 10 10 10,
      keepaspectratio=true]{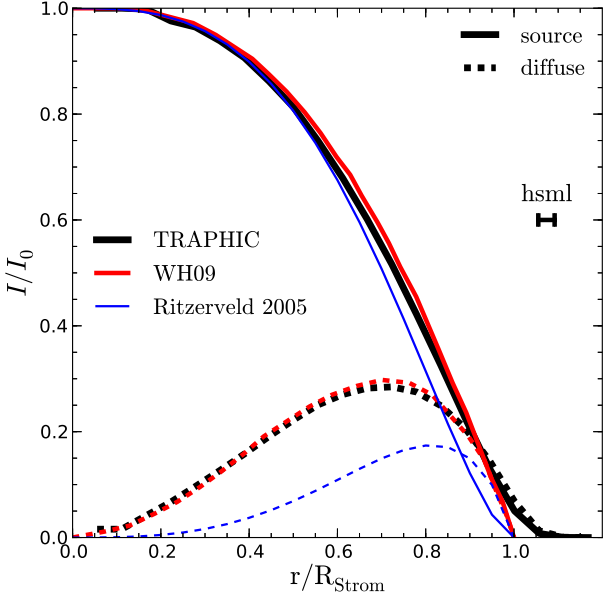}
  \end{center}
  \caption{Equilibrium intensity profile. The profile obtained using
    TRAPHIC (black curves; identical to the corresponding curves in
    the bottom right panel of Figure~\ref{fig1} computed at $t = 4 t_{\rm
      rec}$) is compared to an approximate analytic solution
    \citep[blue curves]{ritzerveld2005} and a numerical solution using
    an equilibrium solver \citep[red curves,][]{williams2009}. Our solution provides an excellent
    match to the numerical reference solution, with only small differences at
    the I-front that can be explained by the lower spatial resolution
    adopted here (the average radius of the SPH kernels is shown by the bar denoted ``hsml''). 
    The analytic solution assumes outward-only transport of the recombination radiation, which explains why it
    underestimates the diffuse-to-source intensity ratio near the source
    and overestimates it near the I-front.}
  \label{fig2}
\end{figure}

The equilibrium state of the HII region has been the standard test of
implementations of recombination radiation in the literature, with
solutions provided both analytically and numerically. In Figure
\ref{fig2}, we compare our equilibrium result (black curves) with two
such works: the analytic approximation of \cite{ritzerveld2005} (blue
curves) and the numerical solution of \cite{williams2009} (red
curves). \cite{ritzerveld2005} provides an analytic approximation for
the equilibrium state of a spherical HII region by assuming that all recombination photons are
transported radially away from the source. This {\it outward approximation}
simplifies the solution of the RT equation greatly as the isotropic
source term of recombination radiation disappears. However, as shown by our simulations, 
the outward approximation results in a radial distribution of the diffuse
intensity that is very different from that in the RT simulation. As already
discussed in \cite{williams2009}, the diffuse intensity near the source is significantly
lower than in the RT solution as all recombination photons are assumed to go
straight towards the edge of the HII region. As a consequence, the
outward-only approximation underestimates the diffuse-to-total intensity
ratio near the source and overestimates it near the I-front.
\par
\cite{williams2009} computed the distribution of diffuse and source intensity 
in stellar HII regions using the method of discrete ordinates \citep[e.g.][]{hummer1963, rubin1968}
which allows for the accurate integration of the RT equation including
the recombination radiation source term. Compared to the
\cite{ritzerveld2005} approach, the \cite{williams2009} calculation shows a
recombination intensity larger by a factor of about 2 near the source,
while still finding it to dominate close to the I-front. The results
of our RT simulation (black curves in Fig. \ref{fig2}) are in
excellent agreement with the results of \cite{williams2009} across the full range of
radii\footnote{The setup of our simulations assumes a hydrogen number density 
  and source luminosity different from those used in \cite{williams2009}. Thus, the comparison
  is only appropriate if the solution is self-similar. In the outward-only approximation discussed in 
  \cite{ritzerveld2005}, this is indeed the case. In their Eqs. 7 and 8, the term in front of
  the brackets on the right-hand side is the total 
number of photons emitted by the source per unit time, and therefore the radiation field ratios 
  depend only on the ratio of the distance from the source and the Stromgren radius. We have not explicitly checked the self-similarity of the numerical solution.}. Small differences exist near the far edge of the I-front, where in our method radiation penetrates slightly further away from the
source. These differences are caused by the limited resolution adopted
in our simulation (the I-front position is slightly pushed forward due
to SPH kernel smoothing).

In conclusion, our new non-equilibrium implementation of recombination radiation in
TRAPHIC yields an equilibrium intensity profile that is in excellent agreement with the
accurate numerical equilibrium solution of \cite{williams2009}. For $r \ll R_{\rm strom}$, direct
source photons dominate the intensity and the ionization balance is close
to that assuming case A. On the other hand, at $r \approx R_{\rm
  strom}$, diffuse recombination photons dominate the intensity and the
ionization balance is close to that assuming case B. Away from these
two limiting recombination regimes, the accurate computation of the 
ionization balance requires an explicit treatment of
the RT of recombination photons. The accurate computation of the 
diffuse-to-direct intensity ratio requires RT for all opacity regimes.

\section{Diffuse radiation in shadowed regions}
\label{sect:shadow}

\begin{figure*}
  \begin{center}
    \includegraphics[width=0.9\textwidth,clip=true, trim=10 10 10 10,
      keepaspectratio=true]{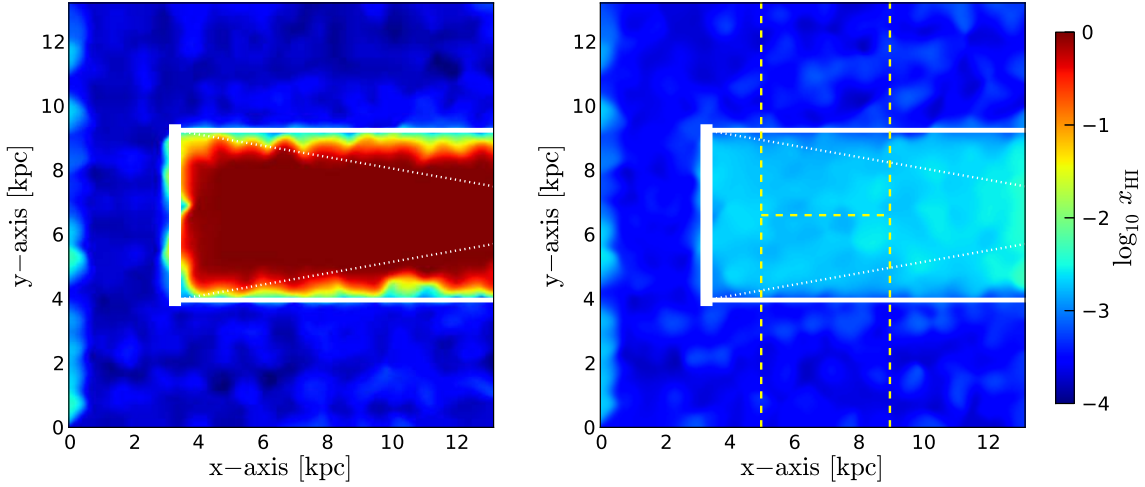}
 \end{center}
  \caption{Shadow region behind an artificial barrier, without ({\it
      left}) and with ({\it right}) the RT of diffuse recombination
    radiation, in the latter case including the contribution of recombination radiation
    originating from outside the simulation volume. Images show the
    neutral fraction field in the $z = L_{\rm box} /2$ plane. The
    artificial barrier is placed at the center of the $x=0.27 L_{\rm
      box}$ plane (thick solid white line). The edges of the shadow in
    the ideal case are indicated by thin solid white lines, while the
    formal angular resolution (i.e. half of the opening angle
    subtended by a transmission cone) is shown using thin white dotted
    lines. The shadow is much sharper than implied by the formal 
    angular resolution, in agreement with the discussion
    in \protect\cite{pawlik2008}. Thin yellow dashed lines mark the region used to compute the
    profiles presented in Figures \ref{fig:shadow_flux}
    and \ref{fig:shadow_T_HI}. Without the RT of recombination
    radiation and adopting the on-the-spot approximation ({\it left}), the shadow remains neutral, while with it
    ({\it right}), it becomes highly ionized. The overlap of the
    shadow with the barrier is due to the smoothing by the
    visualization software. The small HI gradient in the back of the
    shadow region in the right panel is due to the setup not
    accounting for extra recombination photons originating
    from beyond the $x=L_{\rm box}$ side of the box.  }
  \label{fig:shadow_slice}
\end{figure*}

\begin{figure}
  \begin{center}
    \includegraphics[width=0.48\textwidth,clip=true, trim=10 10 10 10,
      keepaspectratio=true]{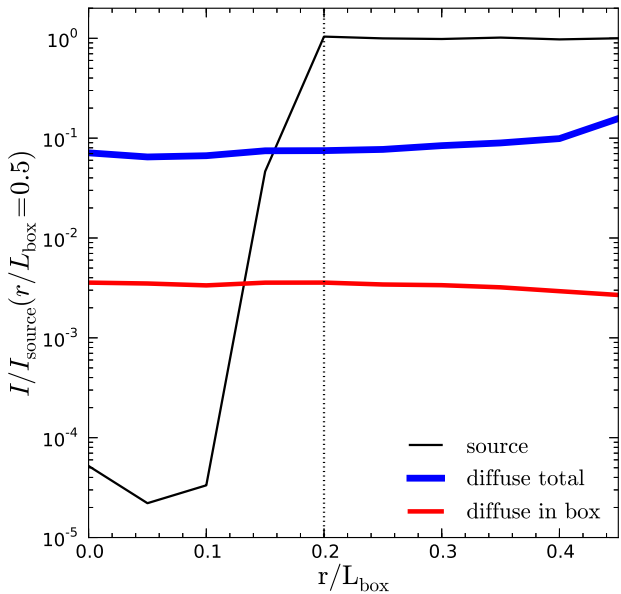}
 \end{center}
  \caption{Intensity of different radiation field components averaged in
    thin cylindrical shells extending radially away from the major axis of
    the shadow. The intensity is measured after ionization equilibrium has
    been reached ($t=3 t_{\rm rec}$).  All values are quoted in units
    of the source radiation intensity averaged in a shell $r=0.5 L_{\rm box}$ away
    from the major axis, $I_{\rm source} (r/L_{\rm box}=0.5)$. The
    source radiation intensity (black solid curve) is constant outside of
    the shadow region (marked by the vertical thin dotted black curve), while
    dropping 4 orders of magnitude inside it. On the other hand, the
    intensity of the diffuse recombination radiation (blue solid curve) is
    nearly independent of distance to the central axis (red dashed curve). The intensity of 
    recombination radiation is dominated by the contributions originating from outside the box.}
  \label{fig:shadow_flux}
\end{figure}

Here, we investigate the impact of recombination radiation on shadows
cast by optically thick objects encountered by an I-front. Such
shadows are often used to demonstrate the geometric precision of a RT
method \citep[e.g.][2013]{iliev2006}. However, most RT methods do not
treat the transport of diffuse recombination radiation which can
penetrate into the shadow region and, in some cases, completely ionize
it. \cite{canto1998} provided an approximate analytic treatment of recombination
radiation in the shadows behind optically thick gas clumps, assuming cylindrical symmetry of the shadow
regions. We apply our implementation to a problem similar to the one studied in
\cite{canto1998}.
\par
We implemented an artificial disk-shaped radiation
barrier and placed it in the path of a plane-parallel I-front entering
the box from one side (see Fig. \ref{fig:boost_diag}). This produces a
cylindrical shadow behind the barrier, i.e., a cylindrical region of
lower ionization compared to the unobstructed surroundings. The
barrier is artificial in the sense that it is not part of the density
field. Instead, for simplicity, it is implemented as a region that
instantly removes all photon packets that pass through it. The barrier
is a disk with radius $r_{\rm barrier}=0.2 L_{\rm box}$ and
is placed parallel to the $x=0$ side of the box at coordinates $(0.27,
0.5, 0.5) \times L_{\rm box}$.
\par
Our setup assumes that the simulation box is fully ionized by an
external source driving a plane-parallel I-front assumed to extend to
infinity in the directions perpendicular to the x-axis, the direction
of propagation. The implementation of this setup requires us to overcome 
a practical problem. In short, in
order for the source to ionize the box, a sufficiently large external
flux must be assumed. The resulting neutral fraction in the non-shadow
region then implies a mean free path for ionizing radiation larger
than the box size. However, the mean free path dictates the amount of
recombination radiation reaching a given point inside the box. When
the mean free path is larger than the box, the contribution to the
recombination radiation field originating in the ionized gas outside
the box has to be added by hand. Here we add this contribution by
placing additional sources at the four sides of the box that are
parallel to the main axis of the shadow region (see below and
Appendix~B). Our setup is idealized but ensures that the shadow region
can become highly ionized by the diffuse radiation. This facilitates the
discussion of the role of heating of the shadow region by recombination photons, which is the focus of this section.
\par
The simulation parameters are as follows: the simulation box is filled
with hydrogen-only gas of uniform density $n_{\rm H}=10^{-3} \, {\rm
  cm^{-3}}$ which is represented using $N_{\rm part} =32^3$ SPH
particles\footnote{The spatial resolution used for the barrier
  shadowing test is intentionally modest to approximate conditions in
  spatially adaptive cosmological simulations. In such simulations,
  most resolution elements are placed in the high density regions that
  cast the shadows, while the shadow regions are sampled at
  significantly lower resolution. However, since the density field is uniform and since 
  TRAPHIC is a photon-conserving RT method, our results are insensitive to resolution.}. The simulation box is cubic with
$L_{\rm box}=13.2 \, \rm kpc$ and is illuminated by the source, a plane-parallel
radiation front\footnote{We use the plane-parallel
  radiation front implementation described in \cite{rahmati2013a}. The
  radiation front is produced by placing virtual particles on a
  regular grid along the box side, with each virtual particle
  representing a photon packet with a direction perpendicular to the
  box side. The number of photons in the packets is set to achieve the
  required flux entering from the simulation side. If the number of
  virtual particles is not very high (here we used $100^2$ particles), SPH particles very close
  to the box side at which the radiation front emerges can be missed by the transmission cones of the
  photons, and these then remain neutral. This causes the noise in the
  neutral fraction field at the left edge of the panels in Figure~\ref{fig:shadow_slice}, 
  but has otherwise no consequences, as the number of photons 
  remains conserved. \label{footnoteifront}} entering the box perpendicular to the $x=0$ side, with 
flux $F_{\rm source}=10^{6} \, \rm photons/s/cm^{2}$. Plane-parallel source
photons, recombination photons emitted inside the box and
recombination photons emitted outside the box are transported in
separate frequency bins, one bin for each radiation component. Following \cite{iliev2006}, we adopt 
a black body spectrum  with temperature $T=10^5 \rm K$, which is characteristic of 
massive metal-free stars, the so-called Population~III. Thus, 
in contrast to the previous section, in which we assumed that
the source photons have energies of $13.6 \eV$, we adopt the 
grey cross-section for absorption of the source photons implied by this 
spectrum, $\langle \sigma_{\rm  HI} \rangle = 1.6 \times 10^{-18} \rm cm^2$, and the associated excess 
photoheating energy $\langle \epsilon_{\rm HI} \rangle = 6.3 \eV$ available to
heat the gas. The RT time step is set to $\Delta t = 0.1 \Myr$, and source photons are injected using an 
emission time step $\Delta t_{\rm em} = \Delta t$. Photons are transported at an angular resolution of 
$N_{\rm TC} = 128$, and merged at an angular resolution of $N_{\rm RC} = 8$. 
\par
The recombination radiation emitted outside the simulation volume is
added at the 4 sides of the box that are parallel to the propagation
direction of the plane-parallel source wave. At each of the 4 sides, a
regular grid of $50^2$ boundary particles is created, each emitting
recombination radiation isotropically. The recombination luminosity of
each boundary particle is determined consistently considering the mean
free path implied by the ionized fraction in front of the barrier. 
Note that we add photons from only 4 out of 6 sides
without adjusting the luminosities of the boundary particles to
compensate for the implied reduction of photons, i.e., we ignore the
recombination radiation emerging from the box sides perpendicular to
the central axis of the shadow. However, the precise recombination radiation intensity is
unimportant to our discussion as long as the shadow is ionized,
because we focus on the role of photoheating in shadows ionized by
recombination radiation. The boundary particles only begin radiating after the I-front driven by the external source has passed them. 
For more details, see Appendix B.

\begin{figure}
  \begin{center}
    \includegraphics[width=0.48\textwidth,clip=true, trim=10 10 10 10]{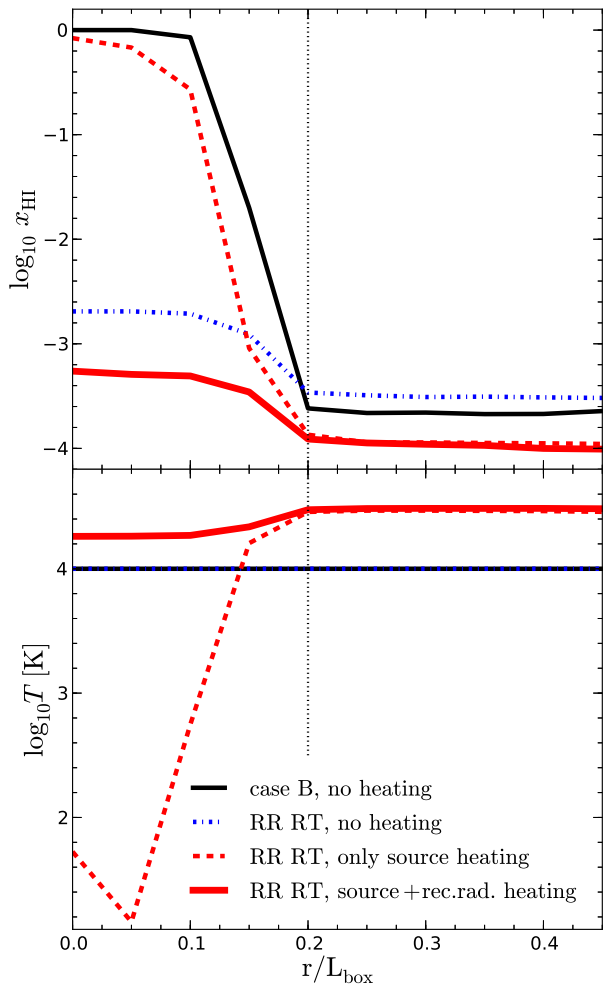}

  \caption{Effect of photoheating. Average HI fraction ({\it top}) and
    temperature ({\it bottom}) in thin cylindrical shells extending
    radially away from the shadow's prime axis at $t = 3 t_{\rm rec}$. In the absence of
    radiation heating (and assuming a constant temperature of
    $10^4\K$), the run adopting the on-the-spot approximation (black solid
    curve) finds a neutral shadow, whereas the run that follows the propagation
    of recombination radiation using RT (blue dot-dashed curve) finds the shadow
    highly ionized. When source photons are allowed to heat
    the gas (red dashed curve), the shadow region remains at the initial temperature 
    of $10^2 K$ or even cools slightly below it (bottom
    panel). The low temperatures imply high recombination rates, 
    leaving the gas mostly neutral (top panel). On the other hand, when also recombination photons are allowed 
    to heat the gas (solid red curve), the shadow region is warm and
    highly ionized. }
  \label{fig:shadow_T_HI}
 \end{center}
\end{figure}

\par
In the following, we present simulations with and without the RT of
recombination radiation. In simulations without RT of recombination
radiation, the on-the-spot approximation is employed, and the
luminosities of the boundary particles are set to zero. We investigate
the role of photoionization heating by carrying out simulations that
ignore photoheating, simulations that include heating only by direct
source photons, and simulations that include heating by both source
and recombination photons. 

\par
Figure \ref{fig:shadow_slice} shows the hydrogen neutral fraction in a
slice through the center (the $z=L_{\rm box} / 2$ plane) of the
simulation box after the equilibrium state has been reached ($t = 3
t_{\rm rec}$, where $t_{\rm rec}=122.4$ Myr, assuming a temperature 
$T = 10^4 \K$). When the on-the-spot approximation is employed,
our simulations find a well-defined neutral shadow region behind
the barrier, with small edge effects that are consistent with the adopted
spatial and angular resolution (left panel). On the other hand, when
the on-the-spot approximation is dropped and recombination radiation is treated using RT, 
the shadow region is highly ionized (right
panel). Both runs ignore the heating of the gas by photoionization,
assuming a constant temperature $T=10^4 \K$ throughout the simulation
box. Apart from the sampling noise due to the finite number of SPH
particles, there are two numerical artifacts visible in Figure
\ref{fig:shadow_slice}. The lower ionization fraction in a thin region
near the $x=0$ plane is caused by some SPH particles being missed by
the photon packets driving the plane-parallel ionization front
(see footnote~\ref{footnoteifront}). The gradient in the HI fraction
in the shadow region ionized by recombination radiation in the right panel
arises because we do not inject recombination radiation originating 
from beyond the $x=L_{\rm box}$ side of the simulation box.
\par
Figure \ref{fig:shadow_flux} shows the intensity of the individual
radiation components. As in Section~\ref{sect:stromgren}, we measured the intensity by 
recording the total number of photons passing through each SPH particle
during a RT time step $\Delta t$ and dividing by this time step. 
However, here we average these numbers in thin cylindrical shells centered on the shadow
region's major axis, i.e., the yellow dashed line parallel to the
shadow in Figure \ref{fig:shadow_slice}. We construct a series of
concentric cylindrical shells starting at this axis and extending radially away from the major axis of the shadow. The radius of the largest shell is
$r=0.5L_{\rm box}$. The height of each cylindrical shell is shown in Figure
\ref{fig:shadow_slice} by two vertical yellow dashed lines. 
The intensities are normalized by dividing by the intensity of source photons 
averaged in a shell $r=0.5 L_{\rm box}$ away from the major axis, $I_{\rm source} (r/L_{\rm box}=0.5)$.
\par
Comparing the intensity of recombination
radiation emitted in the box (red solid curve) and the total diffuse intensity (blue
solid curve) that includes the contribution from photons added to account for the
emission of recombination radiation in ionized gas outside of the box, shows that the
outside-the-box contribution dominates in this setup.
The diffuse recombination radiation intensity is
nearly uniform throughout the box while the source intensity vanishes
nearly completely within the shadow. The small amount of source 
radiation in the shadow is caused by the finite angular
resolution of TRAPHIC and SPH smoothing. The shadow boundaries are
nevertheless much sharper than formally implied by the adopted angular resolution as discussed in \cite{pawlik2008}. Moreover, the small
amount of radiation is not enough to ionize the gas significantly, as
can be seen in the left panel of Fig. \ref{fig:shadow_slice}. 
\par
Figure \ref{fig:shadow_T_HI} shows the effect of photoionization
heating by recombination radiation on the neutral fractions (top) and
temperatures (bottom) in the shadow region. The neutral fractions 
and temperatures were averaged in the same thin cylindrical shells used to compute the
intensity shown in Figure~\ref{fig:shadow_flux}. The two runs that ignore
photoheating, with and without RT of recombination radiation
(blue dot-dashed and black solid curves respectively), assume a
constant temperature of $T=10^4 \K$. Slices through the simulation box
of these two runs were shown in Figure \ref{fig:shadow_slice}. As discussed before, 
the run adopting the on-the-spot approximation yields completely neutral gas in
the shadow region, whereas the other run yields highly ionized gas of 
neutral hydrogen fraction $x_{\rm HI} \sim 10^{-3}$. The difference between
the neutral fractions in the two runs outside of the shadow region is
caused by the difference in the adopted recombination rates: the run
without RT of recombination radiation uses the case B rate whereas
the one with RT of recombination radiation uses case A.
\par
The other two runs presented in Figure \ref{fig:shadow_T_HI} include
photoheating in two variants, one assuming that only direct source
photons heat the gas (red dashed curve) and the other assuming that
both the source and recombination photons heat the gas (solid red
curve). Both of these runs utilize the full RT treatment of
recombination radiation. In our simulations, direct photons heat the
gas outside the shadow to $T_{\rm heat, \, source} \approx 27000 \,
K$. This temperature is used to set the excess energy carried by
recombination photons, $\epsilon_{\rm HI, RR} = k T_{\rm heat, \,
  source} \approx 2.3 \eV$, available to heat the gas following
our discussion in Section~\ref{sect:recrad_method}. For simplicity,
the excess energy is set at the beginning of the simulations to this
value and held constant independent of the actual gas temperatures,
which may deviate slightly from the value quoted above. The initial
temperature in both of these runs is set to $T_{\rm init}=10^2 \rm K$.
\par
The high recombination rate
implied by the low initial temperature causes the shadow region in the
run without  photoheating by recombination radiation (dashed red curve) to
remain substantially neutral, in contrast to the same run without any
photoheating in which we assumed a temperature of $10^4 \K$ (blue
dot-dashed curve). The very low temperatures $< 100 \K$ of the shadow center in
the former run are the result of recombination cooling as the gas is mildly ionized by diffuse photons. 
The cooling is not offset by photoheating, as in this setup recombination radiation is not allowed to heat the gas. As before, 
the small amount of photoionization and photoheating in the shadow region is caused by the finite spatial and angular 
resolution that results in the leakage of source photons beyond the
shadow boundaries.
\par
Setting the excess energy carried 
by recombination photons to zero helps to illustrate
the case in which recombination radiation ionizes the gas but does not
significantly heat it, as may be expected if the recombination radiation spectrum 
is very soft. On the other hand, if we assume that the spectrum of recombination radiation 
is set by the temperature of the gas illuminated by 
a source with a relatively hot blackbody spectrum of $10^5 \K$  (red
solid curve), then the shadow region
is warm and highly ionized. In many practical situations, 
the temperatures and ionized fractions of the shadow will likely lie in between these two 
cases. The run that accounts for photoheating by
both direct and recombination radiation yields a much higher temperature
than the run that only accounts for photoheating by direct photons and, because of the
dependence of the recombination rate on temperature, 
accordingly finds a much lower neutral fraction.
\par
In this section, though the gas in our simulations was initialized to
be neutral and we followed the propagation of the I-front driven by
the external source, we focused on the shadow region in
photo-ionization equilibrium, i.e., we did not discuss the change in
morphology of the shadow during the approach to
equilibrium. \cite{canto1998} showed that, in realistic situations,
depending on the strength of the recombination radiation intensity, in
equilibrium the shadow region will either be fully ionized or contain
a neutral core. However, this does not address the interesting and relevant 
question for how long shadows behind absorbers manage to persist. Additional and less
idealized simulations than those presented here are needed to address
this question. As discussed in \cite{williams2009}, while the diffuse
radiation intensity may be a significant fraction of the intensity of
direct photons, the lateral flux of diffuse photons into the shadow
may be relatively low as the diffuse radiation field may be non-isotropic. Because of these complications, the ability
of recombination radiation to ionize shadows will highly depend on the
specific problem and may be less strong than found here.
\par
In conclusion, we find that the well-defined dark and cold shadows
used to test RT techniques \citep[e.g.][]{iliev2006} do not
necessarily persist when the on-the-spot approximation is abandoned
and the transport of recombination radiation is accurately accounted for. The
exact state of the shadowed region will depend on both the intensity and
the spectrum of both the direct source photons and the diffuse
recombination photons, which may significantly alter the temperatures
and the ionization balance inside the shadow region.

\section{Discussion}
\label{sect:conclude}

We have presented a new, CPU and memory efficient implementation of
ionizing recombination radiation in the TRAPHIC radiative transfer (RT) scheme. The new
implementation improves on the original implementation described in
\cite{pawlik2008} by re-using features already present in the TRAPHIC
scheme. The reception cone directions, used to group photons from
similar directions to control the computational cost of the RT with
TRAPHIC, are re-used as emission directions for recombination
radiation. As in the original implementation, the computational cost is independent
of the number of sources thanks to the merging of photon packets with
similar directions.
\par
We applied our implementation to two problems. First, in Section
\ref{sect:stromgren}, we discussed the role of recombination radiation
in the evolution of a spherical HII region driven by a central
stellar source. We showed, in Figure \ref{fig1}, that, as expected, the widely used
case A and case B approximations for the effect of recombination
radiation on the ionization balance are only accurate in, respectively, the optically thin and
thick regimes and that a detailed treatment of the transport of recombination
radiation is required to obtain accurate solutions in more 
general regimes. The accurate computation of the diffuse-to-source intensity ratio 
always necessitates an accurate RT treatment of recombination photons. 
In equilibrium, our results are in excellent agreement with those 
of \cite{williams2009}, who employed a numeric RT scheme
specifically designed to simulate the effects of recombination radiation in 
HII regions in equilibrium. We confirm the conclusion of \cite{williams2009} that the approximate
analytic solution of \cite{ritzerveld2005}, though in qualitative
agreement with the numerical RT solution and hence providing a substantial improvement
on the commonly employed on-the-spot approximation, underestimates the recombination
radiation intensity relative to the source intensity near the source and overestimates it 
near the I-front.

Second, in Section \ref{sect:shadow}, we investigated the shadowing of
ionizing radiation behind optically thick obstacles. In equilibrium,
the RT treatment of recombination radiation leads to a highly ionized
shadow that is also significantly heated. In Figure
\ref{fig:shadow_T_HI}, we showed that it is important to account self-consistently 
for the radiative heating by recombination radiation. If photoheating
is unimportant, cold gas in the shadow remains substantially neutral
even though it is illuminated by the diffuse flux, because low gas temperatures
imply high recombination rates. These results, which are consistent
with those in earlier works such as \cite{canto1998}, suggest
that shadow regions in RT simulations that treat the radiation using the
on-the-spot approximation may not resemble real shadows, except
perhaps for a brief time just after the onset of irradiation.

This paper focused on the numerical implementation and ignored some
physical intricacies associated with recombination radiation. Since
hydrogen recombination radiation is strongly peaked near the Lyman limit, any
redshifting of the spectrum would make photons unable to ionize
hydrogen. If redshifting is caused by cosmological expansion alone, a
recombination photon would have
to travel $\sim 10^{2} (T/10^{4} \K) (1+z)^{-3/2}$ proper kpc to fall
below the Lyman limit, which at $z = 0$ is much larger than the scales of
interest here. However, redshifting (and blueshifting) may become
important at much smaller scales at higher redshifts or 
if peculiar velocities are non-negligible. We have also ignored recombination radiation emitted by helium. 
The additional complicating factor when helium is included, is 
that helium recombinations can ionize both hydrogen and helium,
requiring special care even when the on-the-spot approximation is used
\citep{cantalupo2011, friedrich2012}. We further ignored effects of spectral
hardening and the absorption of photons by helium and the associated
photoheating, both of which can be important (e.g.,
\citealp{maselli2009}; \citealp{pawlik2011}; \citealp{ciardi2012})

The new implementation of recombination radiation presented in this
paper has already been used by \cite{rahmati2013a,rahmati2013b} to
investigate the effect of ionizing radiation on neutral hydrogen
absorbers in the post-reionization era. There, we demonstrated that
the presence of recombination radiation smooths the transition between
ionized and self-shielded neutral regions. It can also raise the
photoionization rate within formally self-shielded regions by several orders of
magnitude. This result is a good example of the role of recombination
radiation in most common RT problems, usually not dominant but capable of
significantly changing the behaviour of the whole system. Thanks to
its efficiency, our new implementation enables us to dispense with
approximate treatments of the recombination radiation if appropriate.

\section*{Acknowledgments}

The simulations presented here were run on the Cosmology Machine at
the Institute for Computational Cosmology in Durham (which is part of
the DiRAC Facility jointly funded by STFC, the Large Facilities
Capital Fund of BIS, and Durham University) as part of the Virgo
Consortium research programme. This work was sponsored with financial
support from the Netherlands Organization for Scientific Research
(NWO), also through a VIDI grant and an NWO open competition grant. We
also benefited from funding from NOVA, from the European Research
Council under the European Unions Seventh Framework Programme
(FP7/2007-2013) / ERC Grant agreement 278594-GasAroundGalaxies and
from the Marie Curie Training Network CosmoComp
(PITN-GA-2009-238356). AHP receives funding from the European Union's
Seventh Framework Programme (FP7/2007-2013) under grant agreement
number 301096-proFeSsoR.

\bibliographystyle{mn2e} 

\begin{thebibliography}{99}

\bibitem[\protect\citeauthoryear{{Altay}, {Theuns}, {Schaye}, {Booth} \& {Dalla
  Vecchia}}{Altay et~al.}{2013}]{altay2013}
{Altay} G.,  {Theuns} T.,  {Schaye} J.,  {Booth} C.~M.,    {Dalla Vecchia} C.,
  2013, MNRAS, doi:10.1093/mnras/stt1765

\bibitem[\protect\citeauthoryear{Altay, Theuns, Schaye, Crighton \&
  Dalla~Vecchia}{Altay et~al.}{2011}]{altay2011}
Altay G.,  Theuns T.,  Schaye J.,  Crighton N. H.~M.,    Dalla~Vecchia C.,
  2011, ApJ Letters, 737, L37

\bibitem[\protect\citeauthoryear{Aubert 
\& Teyssier}{2010}]{aubert2010} Aubert D., Teyssier R., 2010, ApJ, 724, 244

\bibitem[\protect\citeauthoryear{Barkana \& Loeb}{Barkana \&
  Loeb}{2001}]{barkana2001}
Barkana R.,  Loeb A.,  2001, Physics Reports, 349, 125

\bibitem[\protect\citeauthoryear{Bolton, Meiksin, 
\& White}{2004}]{bolton2004} Bolton J., Meiksin A., White M., 2004, MNRAS, 348, L43

\bibitem[\protect\citeauthoryear{Cantalupo \& Porciani}{Cantalupo \&
  Porciani}{2011}]{cantalupo2011}
Cantalupo S.,  Porciani C.,  2011, MNRAS, 411, 1678

\bibitem[\protect\citeauthoryear{Canto, Raga, Steffen \& Shapiro}{Canto
  et~al.}{1998}]{canto1998}
Canto J.,  Raga A.,  Steffen W.,    Shapiro P.,  1998, ApJ, 502, 695

\bibitem[\protect\citeauthoryear{Ciardi \& Ferrara}{Ciardi \&
  Ferrara}{2005}]{ciardi2005}
Ciardi B.,  Ferrara A.,  2005, Space Science Reviews, 116, 625

\bibitem[\protect\citeauthoryear{Ciardi, Ferrara, Marri \& Raimondo}{Ciardi
  et~al.}{2001}]{ciardi2001}
Ciardi B.,  Ferrara A.,  Marri S.,    Raimondo G.,  2001, MNRAS, 324, 381

\bibitem[\protect\citeauthoryear{Ciardi et al.}{2012}]{ciardi2012} 
Ciardi B., Bolton J.~S., Maselli A., Graziani L., 2012, MNRAS, 423, 558 

\bibitem[\protect\citeauthoryear{Devriendt, Guiderdoni \& Sadat}{Devriendt
  et~al.}{1999}]{devriendt1999}
Devriendt J. E.~G.,  Guiderdoni B.,    Sadat R.,  1999, A\&A, 350, 381

\bibitem[\protect\citeauthoryear{Dijkstra, Haiman, Rees \& Weinberg}{Dijkstra
  et~al.}{2004}]{dijkstra2004}
Dijkstra M.,  Haiman Z.,  Rees M.~J.,    Weinberg D.~H.,  2004, ApJ, 601, 666

\bibitem[\protect\citeauthoryear{Dopita et 
al.}{2011}]{dopita2011} Dopita M.~A., Krauss L.~M., Sutherland R.~S., Kobayashi C., Lineweaver C.~H., 2011, Ap\&SS, 335, 345 
\bibitem[\protect\citeauthoryear{Ercolano 
\& Gritschneder}{2011}]{ercolano2011} Ercolano B., Gritschneder M., 2011, MNRAS, 413, 401 

\bibitem[\protect\citeauthoryear{Faucher-Gigu{\`e}re et 
al.}{2009}]{faucher2009} Faucher-Gigu{\`e}re C.-A., Lidz A., 
Zaldarriaga M., Hernquist L., 2009, ApJ, 703, 1416 

\bibitem[\protect\citeauthoryear{Ferland, Korista, Verner, Ferguson, Kingdon \&
  Verner}{Ferland et~al.}{1998}]{ferland1998}
Ferland G.~J.,  Korista K.~T.,  Verner D.~A.,  Ferguson J.~W.,  Kingdon J.~B.,
    Verner E.~M.,  1998, Publications of the Astronomical Society of the
  Pacific, 110, 761

\bibitem[\protect\citeauthoryear{Ferland, Peterson, Horne, Welsh \&
  Nahar}{Ferland et~al.}{1992}]{ferland1992}
Ferland G.~J.,  Peterson B.~M.,  Horne K.,  Welsh W.~F.,    Nahar S.~N.,  1992,
  ApJ, 387, 95

\bibitem[\protect\citeauthoryear{Finlator, Dav{\'e}, 
{\&Ouml}zel}{2011}]{finlator2011} Finlator K., Dav{\'e} R., {\"O}zel F., 2011, ApJ, 743, 169 

\bibitem[\protect\citeauthoryear{{Friedrich}, {Mellema}, {Iliev} \&
  {Shapiro}}{{Friedrich} et~al.}{2012}]{friedrich2012}
{Friedrich} M.~M.,  {Mellema} G.,  {Iliev} I.~T.,    {Shapiro} P.~R.,  2012,
  \mnras, 421, 2232

\bibitem[\protect\citeauthoryear{Furlanetto, Oh, 
\& Briggs}{2006}]{furlanetto2006} Furlanetto S.~R., Oh S.~P., Briggs F.~H., 2006, PhR, 433, 181

\bibitem[\protect\citeauthoryear{Gnedin}{Gnedin}{2000}]{gnedin2000}
Gnedin N.~Y.,  2000, ApJ, 542, 535

\bibitem[\protect\citeauthoryear{{Gritschneder}, {Naab}, {Burkert}, {Walch},
  {Heitsch} \& {Wetzstein}}{{Gritschneder} et~al.}{2009}]{gritschneder2009}
{Gritschneder} M.,  {Naab} T.,  {Burkert} A.,  {Walch} S.,  {Heitsch} F.,
  {Wetzstein} M.,  2009, \mnras, 393, 21

\bibitem[\protect\citeauthoryear{Gritschneder et 
al.}{2010}]{gritschneder2010} Gritschneder M., Burkert A., Naab T., 
Walch S., 2010, ApJ, 723, 971 

\bibitem[\protect\citeauthoryear{Groves, Dopita, Sutherland, Kewley, Fischera,
  Leitherer, Brandl \& van Breugel}{Groves et~al.}{2008}]{groves2008}
Groves B.,  Dopita M.~A.,  Sutherland R.~S.,  Kewley L.~J.,  Fischera J.,
  Leitherer C.,  Brandl B.,    van Breugel W.,  2008, ApJS, 176, 438

\bibitem[\protect\citeauthoryear{Hasegawa 
\& Umemura}{2010}]{hasegawa2010} Hasegawa K., Umemura M., 2010, MNRAS, 407, 2632 

\bibitem[\protect\citeauthoryear{Hasegawa 
\& Semelin}{2013}]{hasegawa2013} Hasegawa K., Semelin B., 2013, MNRAS, 428, 154

\bibitem[\protect\citeauthoryear{Haworth 
\& Harries}{2012}]{haworth2012} Haworth T.~J., Harries T.~J., 2012, MNRAS, 420, 562 

\bibitem[\protect\citeauthoryear{Hui \& Gnedin}{Hui \& Gnedin}{1997}]{hui1997}
Hui L.,  Gnedin N.~Y.,  1997, MNRAS, 292, 27

\bibitem[\protect\citeauthoryear{Hummer \& Seaton}{Hummer \&
  Seaton}{1963}]{hummer1963}
Hummer D.~G.,  Seaton M.~J.,  1963, MNRAS, 125, 437

\bibitem[\protect\citeauthoryear{Iliev, Ciardi, Alvarez, Maselli, Ferrara,
  Gnedin, Mellema, Nakamoto, Norman, Razoumov, Rijkhorst, Ritzerveld, Shapiro,
  Susa, Umemura \& Whalen}{Iliev et~al.}{2006}]{iliev2006}
Iliev I.~T.,  Ciardi B.,  Alvarez M.~A.,  Maselli A.,  Ferrara A.,  Gnedin
  N.~Y.,  Mellema G.,  Nakamoto T.,  Norman M.~L.,  Razoumov A.~O.,  Rijkhorst
  E.-J.,  Ritzerveld J.,  Shapiro P.~R.,  Susa H.,  Umemura M.,    Whalen
  D.~J.,  2006, MNRAS, 371, 1057

\bibitem[\protect\citeauthoryear{Iliev et al.}{2009}]{iliev2009} 
Iliev I.~T., et al., 2009, MNRAS, 400, 1283 

\bibitem[\protect\citeauthoryear{Iliev, Mellema, Shapiro \& Pen}{Iliev
  et~al.}{2007}]{iliev2007}
Iliev I.~T.,  Mellema G.,  Shapiro P.~R.,    Pen U.-L.,  2007, MNRAS, 376, 534

\bibitem[\protect\citeauthoryear{Inoue}{Inoue}{2010}]{inoue2010}
Inoue A.~K.,  2010, MNRAS, 401, 1325

\bibitem[\protect\citeauthoryear{Jonsson, Groves \& Cox}{Jonsson
  et~al.}{2010}]{jonsson2010}
Jonsson P.,  Groves B.~A.,    Cox T.~J.,  2010, MNRAS, 403, 17

\bibitem[\protect\citeauthoryear{Kang 
\& Shapiro}{1992}]{kang1992} Kang H., Shapiro P.~R., 1992, ApJ, 386, 432 

\bibitem[\protect\citeauthoryear{Mackey}{2012}]{mackey2012} Mackey J., 2012, A\&A, 539, A147 

\bibitem[\protect\citeauthoryear{Maselli, Ciardi, 
\& Kanekar}{2009}]{maselli2009} Maselli A., Ciardi B., Kanekar A., 2009, MNRAS, 393, 171

\bibitem[\protect\citeauthoryear{{McQuinn}, Lidz, Zahn, Dutta, Hernquist \&
  Zaldarriaga}{{McQuinn} et~al.}{2007}]{mcquinn2007}
{McQuinn} M.,  Lidz A.,  Zahn O.,  Dutta S.,  Hernquist L.,    Zaldarriaga M.,
  2007, MNRAS, 377, 1043

\bibitem[\protect\citeauthoryear{{McQuinn}, Oh \& Faucher-Giguere}{{McQuinn}
  et~al.}{2011}]{mcquinn2011}
{McQuinn} M.,  Oh S.~P.,    Faucher-Giguere C.-A.,  2011, ApJ, 743, 82

\bibitem[\protect\citeauthoryear{Mellema et 
al.}{2006}]{mellema2006} Mellema G., Arthur S.~J., Henney W.~J., 
Iliev I.~T., Shapiro P.~R., 2006, ApJ, 647, 397 

\bibitem[\protect\citeauthoryear{Mellema et 
al.}{2006}]{mellemab2006} Mellema G., Iliev I.~T., Pen U.-L., 
Shapiro P.~R., 2006, MNRAS, 372, 679 

\bibitem[\protect\citeauthoryear{{Miralda-Escud{\'e}}}{{Miralda-Escud{\'e}}}{2003}]{miralda-escude2003}
{Miralda-Escud{\'e}} J.,  2003, \apj, 597, 66

\bibitem[\protect\citeauthoryear{Motoyama, Umemoto \& Shang}{Motoyama
  et~al.}{2007}]{motoyama2007}
Motoyama K.,  Umemoto T.,    Shang H.,  2007, A\&A, 467,  657

\bibitem[\protect\citeauthoryear{Norman et al.}{2013}]{norman2013} 
Norman M.~L., Reynolds D.~R., So G.~C., Harkness R.~P., 2013, arXiv, 
arXiv:1306.0645 

\bibitem[\protect\citeauthoryear{Okamoto, Gao \& Theuns}{Okamoto
  et~al.}{2008}]{okamoto2008}
Okamoto T.,  Gao L.,    Theuns T.,  2008, MNRAS, 390, 920

\bibitem[\protect\citeauthoryear{Osterbrock}{Osterbrock}{1989}]{osterbrock1989}
Osterbrock D.~E.,  1989, Astrophysics of gaseous nebulae and active galactic
  nuclei

\bibitem[\protect\citeauthoryear{Pavlakis, Williams, Dyson, Falle \&
  Hartquist}{Pavlakis et~al.}{2001}]{pavlakis2001}
Pavlakis K.~G.,  Williams R. J.~R.,  Dyson J.~E.,  Falle S. A. E.~G.,
  Hartquist T.~W.,  2001, A\&A, 369, 263

\bibitem[\protect\citeauthoryear{Pawlik, Milosavljevi{\'c} \& Bromm}{Pawlik
  et~al.}{2013}]{pawlik2013}
Pawlik A.~H.,  Milosavljevi{\'c} M.,    Bromm V.,  2013, ApJ, 767, 59


\bibitem[\protect\citeauthoryear{Pawlik \& Schaye}{Pawlik \&
  Schaye}{2008}]{pawlik2008}
Pawlik A.~H.,  Schaye J.,  2008, MNRAS, 389, 651

\bibitem[\protect\citeauthoryear{{Pawlik} \& {Schaye}}{{Pawlik} \&
  {Schaye}}{2009}]{pawlik2009}
{Pawlik} A.~H.,  {Schaye} J.,  2009, \mnras, 396, L46

\bibitem[\protect\citeauthoryear{Pawlik \& Schaye}{Pawlik \&
  Schaye}{2011}]{pawlik2011}
Pawlik A.~H.,  Schaye J.,  2011, MNRAS, 412, 1943

\bibitem[\protect\citeauthoryear{Petkova \& Springel}{Petkova \&
  Springel}{2011}]{petkova2011}
Petkova M.,  Springel V.,  2011, MNRAS, 412, 935

\bibitem[\protect\citeauthoryear{Raga et al.}{1999}]{raga1999} 
Raga A.~C., Mellema G., Arthur S.~J., Binette L., Ferruit P., Steffen W., 
1999, RMxAA, 35, 123 

\bibitem[\protect\citeauthoryear{Raga, Henney, Vasconcelos, Cerqueira, Esquivel
  \& Rodriguez-Gonzalez}{Raga et~al.}{2009}]{raga2009}
Raga A.~C.,  Henney W.,  Vasconcelos J.,  Cerqueira A.,  Esquivel A.,
  Rodriguez-Gonzalez A.,  2009, MNRAS, 392, 964

\bibitem[\protect\citeauthoryear{{Rahmati}, {Pawlik}, {Rai\v{c}evi\'{c}} \&
  {Schaye}}{Rahmati et~al.}{2013a}]{rahmati2013a}
{Rahmati} A.,  {Pawlik} A.~H.,  {Rai\v{c}evi\'{c}} M.,    {Schaye} J.,  2013a,
  \mnras, 430, 2427

\bibitem[\protect\citeauthoryear{Rahmati, Schaye, Pawlik \&
  Rai\v{c}evi\'{c}}{Rahmati et~al.}{2013b}]{rahmati2013b}
Rahmati A.,  Schaye J.,  Pawlik A.~H.,    Rai\v{c}evi\'{c} M.,  2013b, MNRAS, 431, 2261

\bibitem[\protect\citeauthoryear{Razoumov \& Scott}{Razoumov \&
  Scott}{1999}]{razoumov1999}
Razoumov A.~O.,  Scott D.,  1999, MNRAS, 309, 287

\bibitem[\protect\citeauthoryear{Ritzerveld}{Ritzerveld}{2005}]{ritzerveld2005}
Ritzerveld J.,  2005, A\&A, 439, L23

\bibitem[\protect\citeauthoryear{{Rosdahl}, {Blaizot}, {Aubert}, {Stranex} \&
  {Teyssier}}{{Rosdahl} et~al.}{2013}]{rosdahl2013}
{Rosdahl} J.,  {Blaizot} J.,  {Aubert} D.,  {Stranex} T.,    {Teyssier} R.,
  2013, MNRAS, doi:10.1093/mnras/stt1722

\bibitem[\protect\citeauthoryear{Rubin}{Rubin}{1968}]{rubin1968}
Rubin R.~H.,  1968, ApJ, 153, 761

\bibitem[\protect\citeauthoryear{Rybicki \& Lightman}{Rybicki \&
  Lightman}{1986}]{rybicki1986}
Rybicki G.~B.,  Lightman A.~P.,  1986, Radiative Processes in Astrophysics

\bibitem[\protect\citeauthoryear{Shapiro, Giroux, 
\& Babul}{1994}]{shapiro1994} Shapiro P.~R., Giroux M.~L., Babul A., 1994, ApJ, 427, 25 

\bibitem[\protect\citeauthoryear{Spitzer}{Spitzer}{1978}]{spitzer1978}
Spitzer L.,  1978, Physical processes in the interstellar medium

\bibitem[\protect\citeauthoryear{Springel}{Springel}{2005}]{springel2005}
Springel V.,  2005, MNRAS, 364, 1105

\bibitem[\protect\citeauthoryear{{Str{\"o}mgren}}{{Str{\"o}mgren}}{1939}]{stromgren1939}
{Str{\"o}mgren} B.,  1939, \apj, 89, 526


\bibitem[\protect\citeauthoryear{Susa}{2008}]{susa2008} Susa H., 
2008, ApJ, 684, 226 

\bibitem[\protect\citeauthoryear{{Teyssier}}{{Teyssier}}{2002}]{teyssier2002}
{Teyssier} R.,  2002, \aap, 385, 337

\bibitem[\protect\citeauthoryear{Trac 
\& Cen}{2007}]{trac2007} Trac H., Cen R., 2007, ApJ, 671, 1

\bibitem[\protect\citeauthoryear{Trac \& Gnedin}{Trac \&
  Gnedin}{2009}]{trac2009}
Trac H.,  Gnedin N.~Y.,  2009, 0906.4348

\bibitem[\protect\citeauthoryear{Verner et al.}{1996}]{verner1996} 
Verner D.~A., Ferland G.~J., Korista K.~T., Yakovlev D.~G., 1996, ApJ, 465, 
487 

\bibitem[\protect\citeauthoryear{Walch et al.}{2013}]{walch2013} 
Walch S., Whitworth A.~P., Bisbas T.~G., Wunsch R., Hubber D.~A., 2013, 
MNRAS, 435, 917

\bibitem[\protect\citeauthoryear{Whalen 
\& Norman}{2008}]{whalen2008} Whalen D., Norman M.~L., 2008, ApJ, 673, 664 

\bibitem[\protect\citeauthoryear{Williams}{Williams}{1999}]{williams1999}
Williams R. J.~R.,  1999, MNRAS, 310, 789

\bibitem[\protect\citeauthoryear{Williams \& Henney}{Williams \&
  Henney}{2009}]{williams2009}
Williams R. J.~R.,  Henney W.~J.,  2009, MNRAS, 400, 263

\bibitem[\protect\citeauthoryear{Wise 
\& Abel}{2008}]{wise2008} Wise J.~H., Abel T., 2008, ApJ, 684, 1 

\bibitem[\protect\citeauthoryear{Wyithe, Mould, 
\& Loeb}{2011}]{wyithe2011} Wyithe S., Mould J., Loeb A., 2011, ApJ,
743, 173

\bibitem[\protect\citeauthoryear{Yajima, Choi, 
\& Nagamine}{2012}]{yajima2012} Yajima H., Choi J.-H., Nagamine K., 2012, MNRAS, 427, 2889

\bibitem[\protect\citeauthoryear{Zaroubi}{2013}]{zaroubi2013} 
Zaroubi S., 2013, ASSL, 396, 45 

\end{thebibliography}

\appendix

\section{Convergence of diffuse radiation in the Stromgren sphere problem}
\label{ap:converge}

\begin{figure}
  \begin{center}
    \includegraphics[width=0.48\textwidth,clip=true, trim=10 10 10 10,
      keepaspectratio=true]{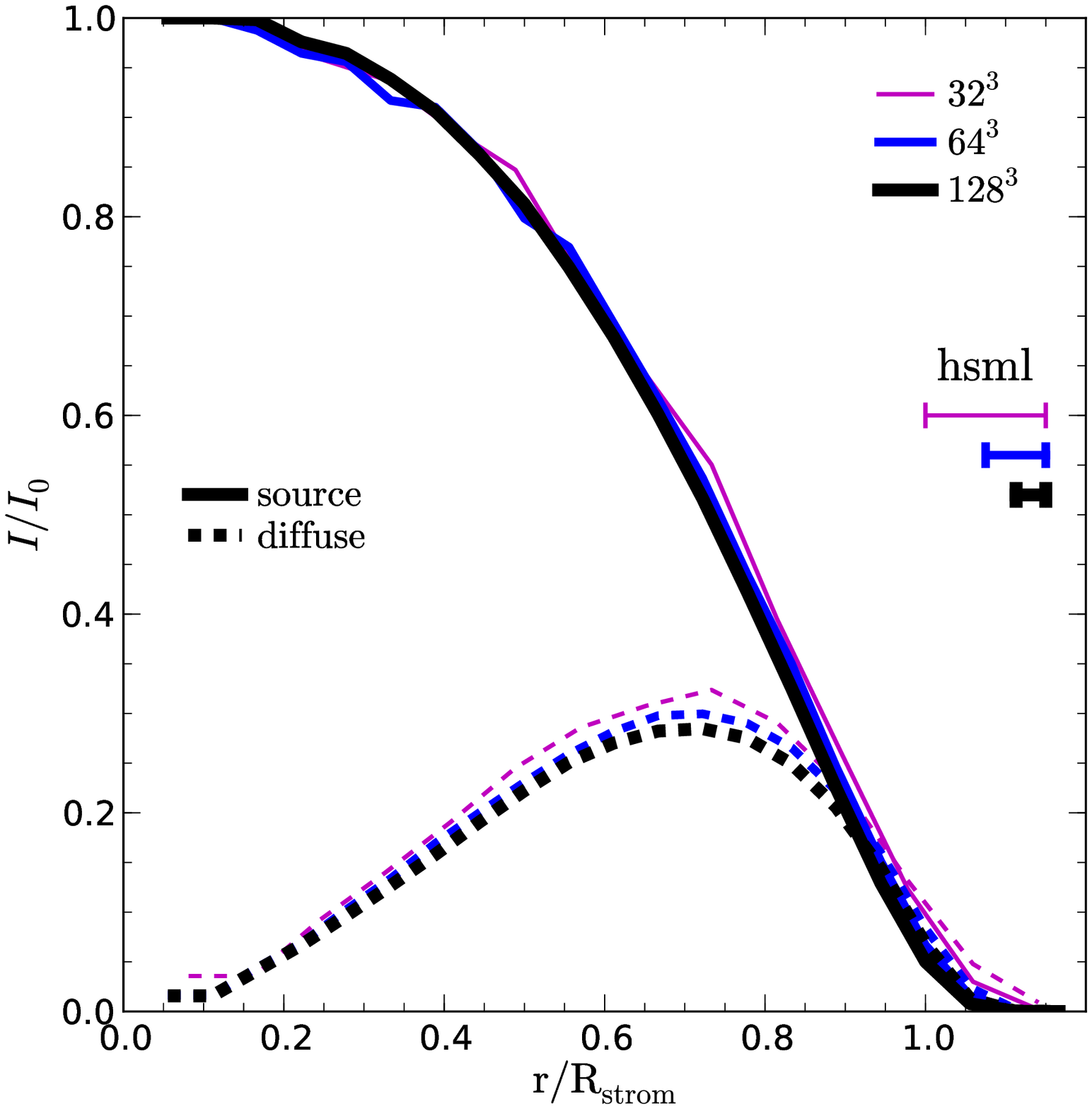}
  \end{center}
  \caption{Convergence of the equilibrium intensity profile with spatial
    resolution. Symbols have the same meaning as in the bottom panels of Figure~\ref{fig1}. A run using $128^3$ SPH
    particles (black curves), i.e., the same number of particles as adopted in the runs discussed in Section~\ref{sect:stromgren} (and identical to the simulation including RT of recombination radiation discussed there), 
    is compared to runs using $32^3$ (magenta curves)
    and $64^3$ (blue curves) particles. Lines to the right, marked
    ``hsml'', show the spatial resolution of each run, i.e. the
    average smoothing kernel of the particles. The runs using $128^3$ and $64^3$ particles are
    converged with respect to spatial resolution.}
  \label{fig:converge_resolution}
\end{figure}

\begin{figure}
  \begin{center}
    \includegraphics[width=0.48\textwidth,clip=true, trim=10 10 10 10,
      keepaspectratio=true]{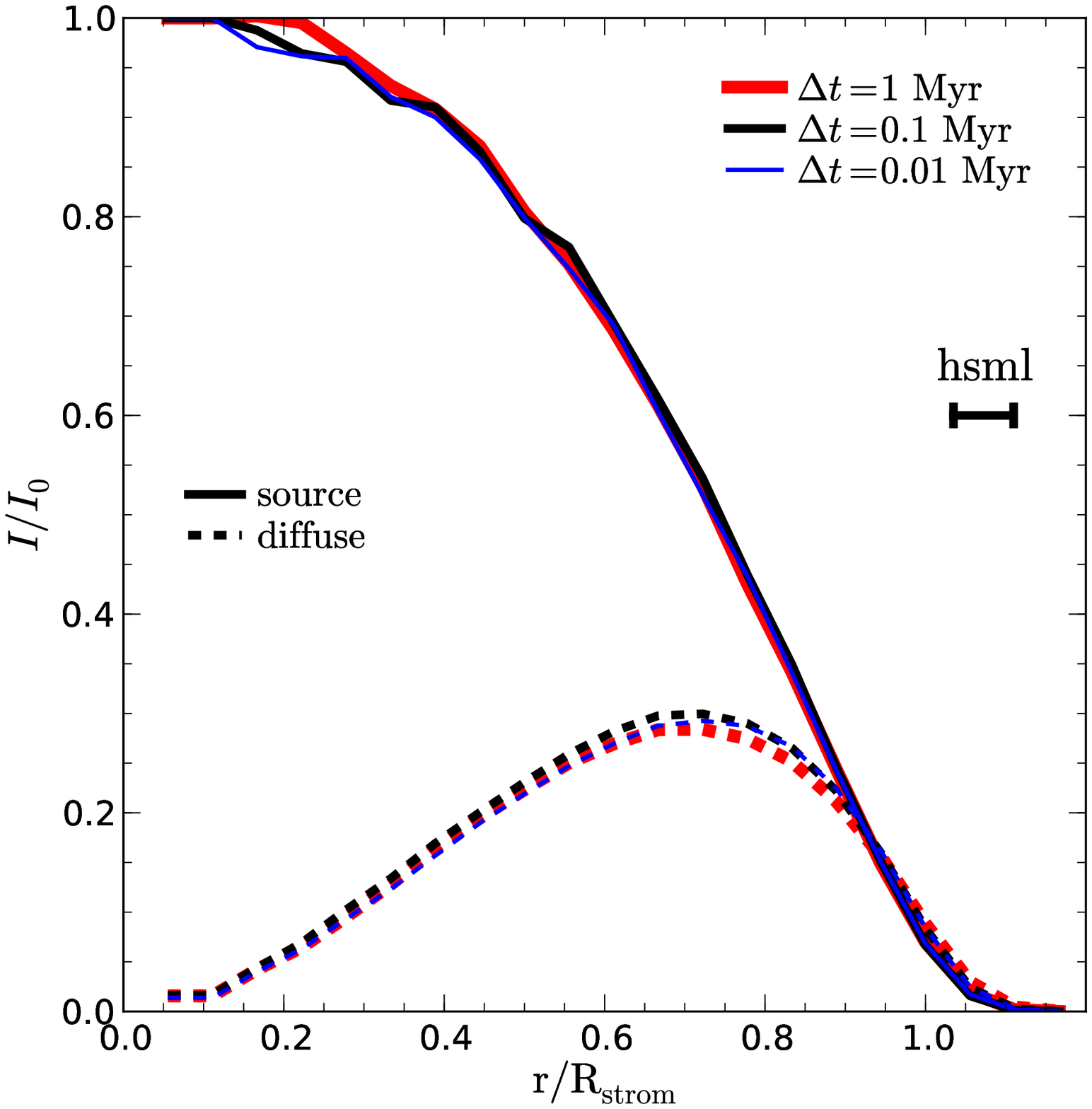}
  \end{center}
  \caption{Convergence of the equilibrium intensity profile with the RT
    time step. Symbols have the same meaning as in the bottom panels of Figure~\ref{fig1}. A run using $\Delta t = 0.1 \, \rm Myr$ (black curves), i.e., the time step 
    adopted in the runs discussed in Section~\ref{sect:stromgren}, is compared to runs with
    10 times longer (red curves) and shorter (blue curves) time
    steps. In all runs, $\Delta t_{\rm em} = 0.1 \Delta t$. The
    run using $\Delta t = 0.1 \, \rm Myr$ is converged.}
  \label{fig:converge_timestep}
\end{figure}

In this appendix, we discuss the convergence of our simulations of the
expansion of an HII region including recombination radiation
(Fig. \ref{fig2}). We present convergence tests at $t = 4t_{\rm
  rec}$, i.e., in ionization equilibrium, which is the relevant case to compare
our method to solutions discussed in the literature (Section
\ref{sect:stromgren}). We note that the convergence at earlier times
is equally good.

Figure~\ref{fig:converge_resolution} compares the run discussed in
Section~\ref{sect:stromgren} and carried out using $128^3$ (black
curves; identical to the simulation including RT of recombination radiation shown 
in Figs.~\ref{fig1} and \ref{fig2}) with two lower-resolution runs employing 8 (magenta curves)
and 64 (blue curves) times less particles to discretize the density
field. The intensity profiles in the runs using $128^3$ and $64^3$ are
nearly identical.  The lowest resolution simulation using $32^3$
particles yields an intensity that is significant out to slightly larger
radii than in the higher resolution runs. However, the difference is
within the size of the resolution element (see the smoothing length
measures on the right hand side of the figure).
\par
For the following convergence tests, a spatial resolution of $64^3$
particles is adopted which, as we have just shown, yields converged
results in terms of spatial resolution. In Figure
\ref{fig:converge_timestep}, we compare a run using the time step
used in Section~\ref{sect:stromgren} (black
curve) with runs employing one order of
magnitude shorter and longer time steps, $\Delta t = 0.01 \rm \, Myr$
and $\Delta t = 1 \rm \, Myr$ respectively. In all cases, the emission
time step is $\Delta t_{\rm em} = 0.1 \Delta t$. All three runs give
nearly identical results.

\begin{figure}
  \begin{center}
    \includegraphics[width=0.48\textwidth,clip=true, trim=10 10 10 10,
      keepaspectratio=true]{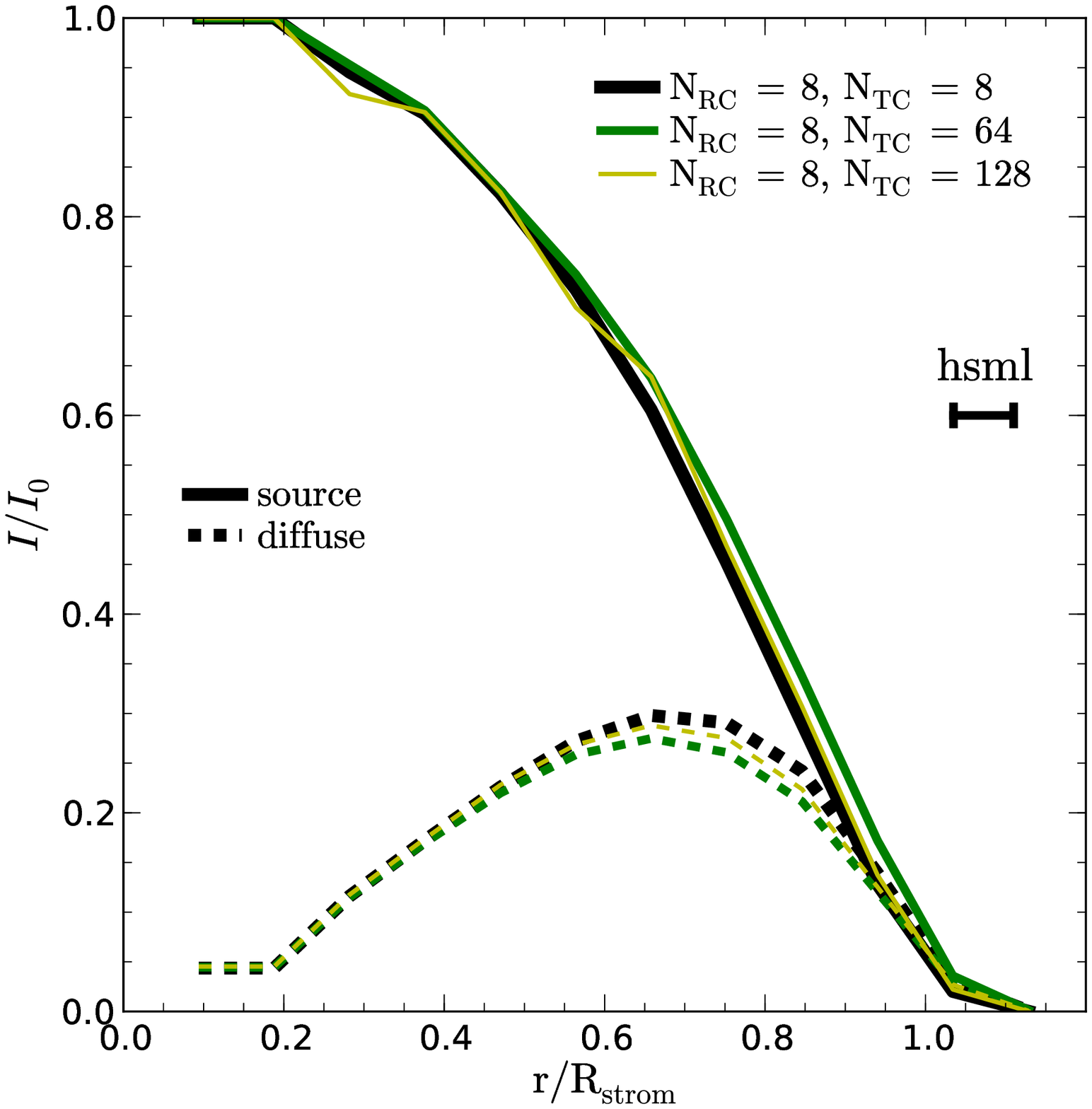}
  \end{center}
  \caption{Convergence of the equilibrium intensity profile with angular
    resolution, expressed in terms of the number of transmission cones
    $N_{\rm TC}$. Symbols have the same meaning as in the bottom panels of Figure~\ref{fig1}. A run using $N_{\rm
      TC}=8$, i.e., the same angular resolution as employed in the runs discussed in Section~\ref{sect:stromgren},
    is compared with two runs using higher angular
    resolution, $N_{\rm TC}=64$ (green curves) and $N_{\rm TC}=128$
    (yellow curves). All runs use $N_{\rm EC} = N_{\rm RC} =8$ cones
    to emit stellar and recombination photons,
    respectively. The small differences between the runs result 
    because we increased the angular resolution without increasing
    the angular sampling.}
  \label{fig:converge_Ncones}
\end{figure}

Finally, we investigate the impact of varying
the angular resolution at which the RT is carried out. In Figure~\ref{fig:converge_Ncones}, 
we decrease the solid angle subtended by
each transmission cone from 
$4 \pi / 8$ (black) to $4 \pi / 64$ (green) and $4
\pi / 128$ (yellow). Most of the small intensity differences seen in the runs at different
angular resolutions are caused by the increasingly poorer angular
sampling, which is because we kept $N_{\rm EC} = N_{\rm RC} = 8$ fixed
while increasing the angular resolution (i.e., $N_{\rm TC}$). Indeed,
additional tests have shown that the small differences seen in Figure
\ref{fig:converge_Ncones} disappear when the angular sampling is
increased in step with the angular resolution, by increasing $N_{\rm
  EC}$ and $N_{\rm RC}$, as expected given the spherical symmetry of the problem.

\section{Adding the missing recombination radiation intensity to the shadowing test setup}
\label{ap:boost}

In our simulations presented in Section~\ref{sect:shadow}, we assume
that an infinitely extended plane-parallel front of ionizing radiation
thought to originate from a source outside the box enters the box from
one side and moves to ionize the simulation box, leaving a shadow
behind the opaque obstacle present in the box. The source intensity
driving the plane-parallel front determines the ionization balance 
inside the box, setting the mean free path of the ionizing photons and the distance 
out to which an I-front is driven. Because we require the entire box to be ionized,
both the distance to the I-front and the mean free path must be larger
than (or at least equal to) the length of the box. Here we show that
this means that recombination radiation emitted from ionized gas
outside the box contributes to the ionization balance inside the box
(assuming that the cross-sections for the absorption of stellar and
recombination photons and hence the mean free paths are similar), and describe how we account for
this contribution.
\par
Consider a simulation box containing  
hydrogen-only gas with uniform number density $\nh$, illuminated by a
plane-parallel radiation front of flux $F$, as in Section
\ref{sect:shadow}. Equilibrium between ionizations and recombinations
implies
\begin{equation}
F A = \alpha_{B} n_{\rm H}^{2} A L_{\rm strom},
\end{equation}
where $A$ is the surface area of the box side through which the plane
parallel radiation front of flux $F$ is entering the box and $L_{\rm strom}$, the Stromgren distance, is the
distance that the I-front will penetrate \citep[e.g.,][]{gritschneder2009}, 
given by
\begin{equation}
\label{eq:strom_dist}
L_{\rm strom} = \frac{F}{\alpha_{B}n_{\rm H}^{2}}.
\end{equation}
In our simulations, we require the Stromgren distance to be larger than the box side length
$L_{\rm box}$. We achieve this by setting the incoming flux $F$ such
that Eq. \ref{eq:strom_dist} gives $L_{\rm strom} > L_{\rm box}$. In
ionization equilibrium, the choice of the flux $F$ then implies a
value for the ionized fraction of the gas determined using
\begin{equation}
\label{eq:ion_balance}
\Gamma_{\rm ion} n_{\rm H} (1-x_{\rm HII}) = \alpha_{B} n_{\rm H}^{2}
x_{\rm HII}^{2},
\end{equation}
where  $x_{\rm HII} =
1-x_{\rm HI}$ is the ionized hydrogen fraction, $x_{\rm HI}$ the neutral hydrogen fraction,
$\Gamma_{\rm ion} = \sigma F$ the photoionization rate
of the incoming photons, and $\sigma$ the photoionization
cross-section. For simplicity, but without loss of generality, we 
consider the case of monochromatic radiation.
\par
The recombination radiation emissivity per unit volume of gas with
ionized fraction $x_{\rm HII}$ computed using Eq. \ref{eq:ion_balance} is
\begin{equation}
\label{eq:recrad_emissivity}
E_{\rm RR} = \alpha_{1} n_{\rm H}^2 x_{\rm HII}^2,
\end{equation}
while the mean free path of recombination radiation through the same
gas is
\begin{equation}
\label{eq:mfp}
\lambda_{\rm mfp} = (\sigma n_{\rm H} (1-x_{\rm HII}))^{-1}.
\end{equation} 

\begin{figure}
  \begin{center}
    \includegraphics[width=0.48\textwidth,clip=true, trim=10 10 10 10,
      keepaspectratio=true]{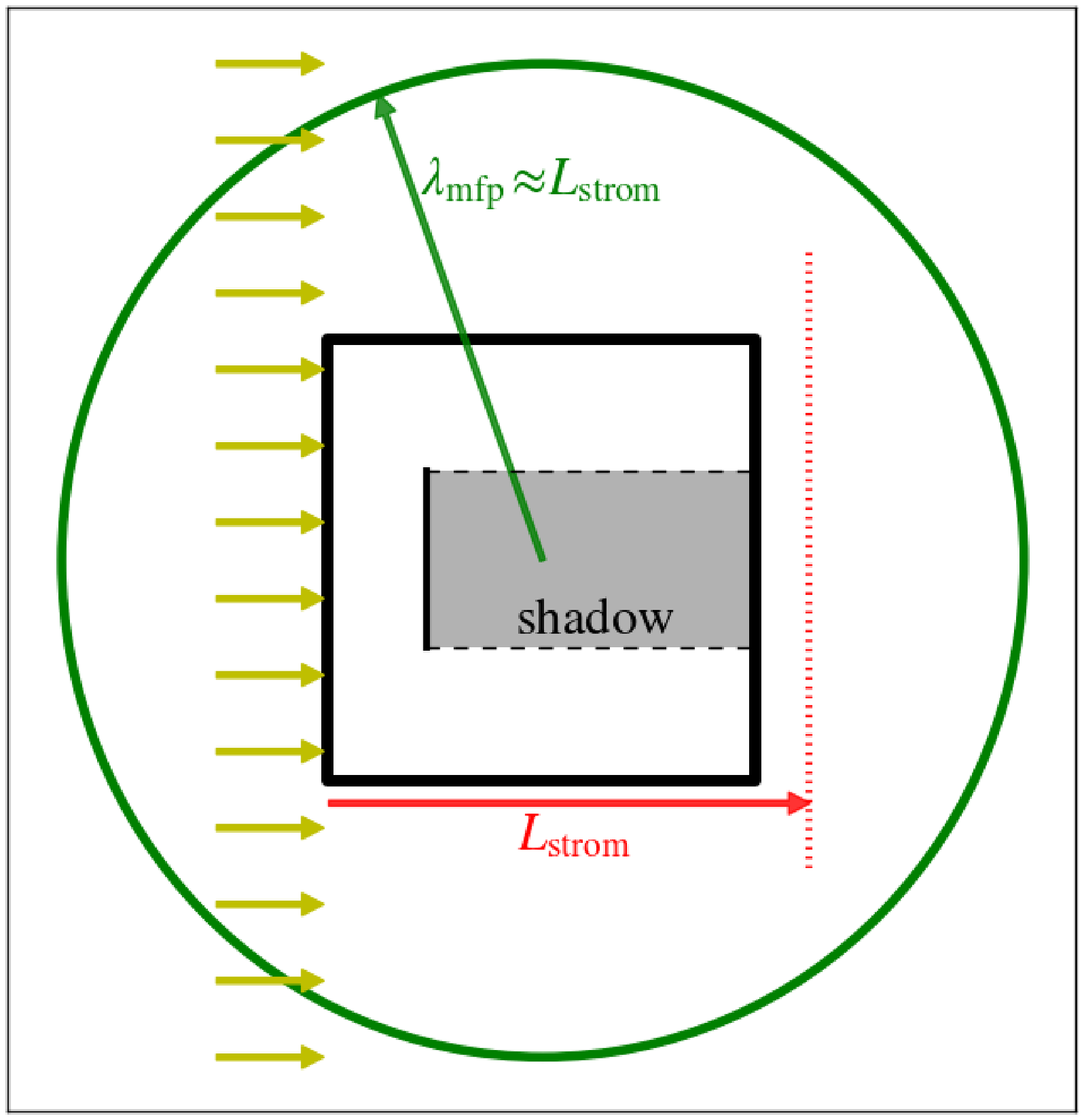}
  \end{center}
  \caption{Diagram describing the issue of the shadowing test setup
    where some of the sources of recombination radiation are not
    encompassed by the simulation box. A plane-parallel radiation front
    assumed to extend infinitely in the directions perpendicular the direction of propagation (yellow arrows), 
    enters the simulation box (thick solid black cube)
    from one side, causing the barrier in the box (thin
    solid black line) to cast a shadow behind it (gray
    shaded area bordered by dashed black lines). In order for the
    shadow to extend to the edge of the box, the Stromgren distance
    associated with the incoming plane parallel flux, $L_{\rm strom}$
    (red dotted line), must be larger than the box, $L_{\rm box}$. The
    plane-parallel flux intensity also determines the ionized fraction
    in the gas around the shadow, which in turn defines the mean free
    path of the ionizing recombination photons, $\lambda_{\rm mfp}$ (green solid
    arrow). Since we require that $L_{\rm strom} \gtrsim
    L_{\rm box}$, the region from which the recombination radiation
    can reach the shadow extends beyond the size of the simulation box
    (shown by the solid green circle), because $L_{\rm strom} \approx
    \lambda_{\rm mfp}$. In our simulations, the recombination radiation originating
    from outside the box is added by hand.}
  \label{fig:boost_diag}
\end{figure}

The key issue is that, not surprisingly, $L_{\rm strom}$ and $\lambda_{\rm mfp}$ 
take very similar values. E.g., using the values for $F$ and $n_{\rm
  H}$ assumed in Section \ref{sect:shadow} in Eqs. \ref{eq:strom_dist},
\ref{eq:ion_balance} and \ref{eq:mfp} yields $L_{\rm strom} =
1251.25 \, \rm kpc$ and $\lambda_{\rm mfp} = 1251.37 \, \rm kpc$,
i.e., $L_{\rm strom} \approx \lambda_{\rm mfp}$. This similarity causes a practical 
problem since in order for a point in the shadowed region to see the full
intensity of the recombination radiation emitted by the surrounding gas, the
computation box must be large enough to include a sphere of radius
$\lambda_{\rm mfp}$ around the center of the
shadow. Our setup assumes that the whole simulation
box is ionized by the source flux, which implies $L_{\rm strom} >
L_{\rm box}$. However, because $\lambda_{\rm mfp} \approx L_{\rm strom}$, it
is not possible for the box of size $L_{\rm box}$ to contain a sphere
with radius $\lambda_{\rm mfp}$. Figure \ref{fig:boost_diag} illustrates
the issue.
 
We account for the contribution to the intensity from recombining gas
outside the box by inserting additional particles positioned at the
four sides of the computational box parallel to the major axis of the
shadow region, whom we then assign an appropriate luminosity. We do
not place such boundary particles at the $x=0$ side in order to have a
strict control of the number of photons entering the box from the
source side. We do not place boundary particles at the $x=L_{\rm box}$ side because the
shadow covers a portion of it. The luminosity of the boundary
particles is set to reproduce the total number of recombination
radiation photons through a surface with area $A$, emitted by the gas with
uniform density $n_{\rm H}$ and ionized fraction given by
Eq. \ref{eq:ion_balance}. The associated intensity is given by the integral
of all recombination radiation emitted by uniformly ionized gas of constant
hydrogen density $n_{\rm H}$ arriving at the centre of a 
sphere with radius $\lambda_{\rm mfp}$,
\begin{equation}
I = 4 \pi \int_{0}^{\lambda_{\rm mfp}} \frac{E_{\rm RR} e^{-\sigma n_{\rm H}  x_{\rm HI} r}}{4 \pi r^2} r^2 {\rm d}r = 0.63 \lambda_{\rm mfp} E_{\rm RR}. 
\end{equation} 
The fraction under the integral is the recombination radiation
intensity coming from a single point at distance $r$ and to obtain the
final solution we used $\lambda_{\rm mfp} = (\sigma n_{\rm H} x_{\rm
  HI})^{-1}$.  The neutral fraction $x_{\rm HI}$ used here is the
neutral fraction found in the gas in front of the barrier in a
simulation that adopts the on-the-spot approximation. The boundary
sources are distributed on a regular grid and switched on only after
being overtaken by the I-front driven by the external source. The
number of boundary sources is chosen to achieve numerical
convergence. In the simulations in Section~\ref{sect:shadow}, we used
$50^2$ boundary sources per side. Note that for simplicity, we do not
adjust the flux of recombination radiation emitted by the boundary
particles at the 4 sides to compensate for the lack of emission from
the 2 remaining sides. This approximation does not affect our
conclusions, because we find that the shadow is highly ionized by
recombination radiation even without such a compensation.

\label{lastpage}

\end{document}